\documentclass[a4paper, notitlepage, 11pt]{article}
\usepackage{geometry}
\fontfamily{times}
\usepackage{setspace,relsize}               
\usepackage{moreverb}                        
\usepackage{url}
\usepackage{hyperref}
\hypersetup{colorlinks=true,citecolor=blue}
\usepackage{amsmath}
\usepackage{mathtools} 
\usepackage{amsthm}
\usepackage{amssymb}
\usepackage{indentfirst}
\usepackage{todonotes}
\usepackage{subfigure}
\usepackage{multirow}
\usepackage[authoryear,round]{natbib}
\usepackage[pdftex]{lscape}
\usepackage[utf8]{inputenc}
\title{\vspace{-9ex}\centering \bf Combining probability distributions: Extending the logarithmic pooling approach
}
\author{
Luiz Max F. de Carvalho$^{a}$, Daniel A. M. Villela$^b$, Flavio Coelho$^a$ \& Leonardo S. Bastos$^b$ \\
a -- School of Applied Mathematics, Get\'ulio Vargas Foundation (FGV)\\
b -- Program for Scientific Computing (PROCC), Oswaldo Cruz Foundation. \\
}
\DeclareMathOperator*{\argmin}{arg\,min}
\DeclareMathOperator*{\argmax}{arg\,max}
\DeclareMathOperator\supp{supp}
\newtheorem{theo}{Theorem}[]

\newtheorem{remark}{Remark}[]
\newtheorem{lemma}{Lemma}[]
\begin{document}
\maketitle

\begin{abstract}
Combining distributions is an important issue in decision theory and Bayesian inference.
Logarithmic pooling is a popular method to aggregate expert opinions by using a set of weights that reflect the reliability of each information source.
However, the resulting pooled distribution depends heavily on set of weights given to each opinion/prior and thus careful consideration must be given to the choice of weights.
In this paper we review and extend the statistical theory of logarithmic pooling, focusing on the assignment of the weights using a hierarchical prior distribution. 
We explore several statistical applications, such as the estimation of survival probabilities, meta-analysis and Bayesian melding of deterministic models of population growth and epidemics.
We show that it is possible learn the weights from data, although identifiability issues may arise for some configurations of priors and data.
Furthermore, we show how the hierarchical approach leads to posterior distributions that are able to accommodate prior-data conflict in complex models.

\end{abstract}

\section{Introduction}
\label{sec:intro}

Combining probability distributions is a topic of general interest, both in the statistical~\citep{West1984, Genest1986A, Genest1986B} and decision theory literatures~\citep{Genest1984,French1985,Pennock1997,Guardoni2002}.
On the theoretical front, studying opinion pooling operators may give important insights on consensus belief formation and group decision making~\citep{West1984,Genest1986B,Guardoni2002}.
Among the various opinion pooling operators proposed in the literature, logarithmic pooling has enjoyed much popularity, mainly due to its many desirable properties such as relative propensity consistency (RPC) and external Bayesianity (EB)~\citep{Genest1986A} -- see below. 
In a practical setting, logarithmic pooling finds use in a wide range of fields, from engineering~\citep{Lind1988,Savchuk1994} to wildlife conservation~\citep{Poole2000} and infectious disease modelling~\citep{Coelho2009}. 

A common situation of interest is combining expert opinions about a quantity of interest $\theta \in \mathbf{\Theta} \subseteq \mathbb{R}^p$ when these opinions can be represented as (proper) probability distributions.
Combining these opinions using logarithmic pooling requires assigning weights to each of the experts, which represent the (relative) reliability of each opinion~\citep{Genest1984,French1985}.
This requirement naturally leads to the question of how to choose the weights in a meaningful way, according to some well-accepted (optimality) criterion.
There are a few proposals in the literature that build methods using different approaches.
One proposal is to maximise the entropy the pooled distribution~\citep{Myung1996}, whereas another one is to minimise Kullback-Leibler (KL) divergence between the pooled distribution and the individual opinions~\citep{Abbas2009} or between the pooled (prior) distribution and the posterior distribution~\citep{Rufo2012A,Rufo2012B}.

While moving away from the problem of arbitrarily assigning the weights, these approaches arrive at single point solutions, similar to point estimates in statistical theory.
While we acknowledge that these approaches have merit, we argue that in many settings it would be desirable to incorporate  information on the relative reliabilities of the experts into the pooling procedure while accommodating uncertainty about the weights.
Moreover, assigning a probability distribution over the weights allows one to obtain a posterior distribution using a Bayesian procedure, which in turn enables learning about the weights from data~\citep{Poole2000}.
Therefore, it makes possible to sequentially update knowledge about the reliability of each expert/source in the face of new data.

In this paper we discuss previous approaches for assigning the weights based on optimality criteria and study assigning hierarchical priors to the weights in order to learn about them from data.
This paper is organised as follows: in Section~\ref{sec:background} we introduce the necessary concepts and notation on logarithmic pooling, as well as some its key properties.
We also prove a new result about log-concavity of the pooled distribution when all distributions are log-concave.
In Section~\ref{sec:weights} we present different approaches to choosing the weights, two methods based on optimality criteria, namely maximising the entropy of the pooled prior and minimising Kullback-Leibler divergence between the pooled distribution and the expert distributions.
In addition we also lay out an approach hierarchical modelling of the weights.
Section~\ref{sec:apps} contains applications of logarithmic pooling to reliability analysis (Sections~\ref{sec:survivalProbs} and~\ref{sec:learning_rate}), meta-analysis (Section~\ref{sec:metaAnalysis}) and Bayesian melding (Section~\ref{sec:melding_apps}).
We conclude with a discussion of our results in light of the statistical literature in Section~\ref{sec:discussion}.

\section{Logarithmic pooling: properties and applications}
\label{sec:background}

In this section we introduce the necessary theory and notation and motivate the use of the logarithmic pooling operator by presenting some of its desirable properties.

First let us define the logarithmic pooling (LP) operator.
Let $\mathbf{F}_{\theta} := \{f_0(\theta), f_1(\theta), \ldots, f_K(\theta)\}$ be a set of (densities of) distributions representing the opinions of $K+1$ experts and let $\boldsymbol\alpha :=\{\alpha_0, \alpha_1, \alpha_2, \ldots, \alpha_K \}$ be the vector of weights, such that $\alpha_i > 0\: \forall i$ and $\sum_{i=0}^K \alpha_i = 1$.
The log-pooled density is
\begin{equation}
\label{eq:logpool}
 \mathcal{LP}(\mathbf{F_\theta}, \boldsymbol\alpha) := \pi(\theta \mid \boldsymbol\alpha) = t(\boldsymbol\alpha) \prod_{i=0}^K f_i(\theta)^{\alpha_i},
\end{equation}
where $t(\boldsymbol\alpha) = \left[ \int_{\boldsymbol\Theta}\prod_{i=0}^K f_i(\theta)^{\alpha_i}\, d\theta \right]^{-1}$.

Logarithmic pooling will only yield proper probability distributions if it is possible to normalise the expression in (\ref{eq:logpool}).
This condition is usually assumed implicitly, without proof.
While \citet{Poole2000} provide a proof for the case of two densities (see Theorem 1 therein),~\cite{Genest1986A} (pg.489) prove the result for a finite number of densities:
\begin{theo}
\label{thm:normalisation}
\textbf{Normalisation~\citep{Genest1986A}}. 
Let $\mathcal{A}$ be a $(K+1)$-dimensional open simplex on $[0,1]$.
For all $\boldsymbol\alpha \in \mathcal{A}$ there exists a constant $t(\boldsymbol\alpha)$ such that $\int_{\boldsymbol\Theta}\pi(\theta \mid \boldsymbol \alpha)\, d\theta = 1$.
\end{theo}
We give a simple proof using H\"{o}lder's inequality in the Appendix.
This result ensures any (finite) number of proper distributions can be combined using the logarithmic pooling operator to yield a normalisable (proper) density.
In addition, log-linear pools enjoy the~\textit{external Bayesianity} property (Remark~\ref{rmk:properties_EB}), which guarantees that whether one combines the expert opinions before or after observing evidence does not affect the resulting pooled distribution.
\begin{remark}
\label{rmk:properties_EB}
 \textbf{External Bayesianity~\citep{Genest1984}}.
 If the expert opinions are given by densities $f_i(\theta)$ and one observes data $x$ such that one can specify a likelihood $l(x \mid \theta)$, combining the set of posteriors $p_i(\theta \mid x) \propto  l(x \mid \theta)f_i(\theta) $ yields the same distribution as combining the densities $f_i$ to obtain a prior $\pi(\theta)$ and then combine it with $l(x \mid \theta)$ to obtain a posterior $p(\theta \mid x) \propto l(x \mid \theta)\pi(\theta)$.
\end{remark}
\begin{proof}
 Combining the posteriors $p_i(\cdot)$ gives
 \begin{align*}
  p^\prime (\theta \mid x, \boldsymbol \alpha) &\propto \prod_{i = 0}^K \left[  l(x \mid \theta)f_i(\theta) \right]^{\alpha_i},\\
  &\propto   l(x \mid \theta) \prod_{i = 0}^K f_i(\theta)^{\alpha_i},\\
  &=  \frac{l(x \mid \theta)\pi(\theta \mid \boldsymbol \alpha)}{m^{\prime}(x)} \equiv   p(\theta \mid x, \boldsymbol \alpha),
 \end{align*}
 where the second line follows from $\sum_{i=0}^K \alpha_i = 1$.
\end{proof}
\cite{Genest1984} show that the logarithmic pooling operator in~(\ref{eq:logpool}) is the~\textbf{only} aggregation (pooling) operator that enjoys external Bayesianity.
Moreover, the logarithmic pooling operator has the relative propensity consistency (RPC) property (Remark~\ref{rmk:properties_RPC}), whereby the pooled opinion preserves relative judgments from the experts.
\begin{remark}
\label{rmk:properties_RPC}
\textbf{Relative propensity consistency~\citep{Genest1984}}.
Taking $\boldsymbol F_{X}$ as a set of expert opinions with support on a space $\mathcal{X}$, define $\boldsymbol \xi = \{\boldsymbol F_{X}, a, b\}$ for arbitrary $a , b \in \mathcal{X}$.
Let $\mathcal{T}$ be a pooling operator and define two functions $U$ and $V$ such that 
\begin{align}
 U(\boldsymbol \xi) &:= \left( \frac{f_0(a)}{f_0(b)}, \frac{f_1(a)}{f_1(b)}, \ldots, \frac{f_K(a)}{f_K(b)} \right)\:\text{and}\\
 V(\boldsymbol \xi) & := \frac{\mathcal{T}_{\boldsymbol F_{X}} (a)}{\mathcal{T}_{\boldsymbol F_{X}} (b)}.
\end{align}
We then say that $\mathcal{T}$ enjoys \textit{relative propensity consistency} (RPC) if and only if
\begin{equation}
 U(\boldsymbol \xi_1) \geq U(\boldsymbol \xi_2) \implies  V(\boldsymbol \xi_1) \geq V(\boldsymbol \xi_2),
\end{equation}
for all $\boldsymbol \xi_1, \boldsymbol \xi_2$.
\end{remark}
We refer the reader to~\cite{Genest1984} for a proof.
Informally, this property says that if all experts consider a particular event $A$ more probable than another event $B$, then the pooled opinion should be consistent with these relative judgments. 
\cite{Genest1984} show that for mild conditions on $\mathcal{X}$, namely $|\mathcal{X}| \geq 3$, the logarithmic pooling operator is the only pooling operator with RPC (see also Lemma~\ref{lem:RPC_representation} in Appendix~\ref{sec:appendix_proofs}).

Another desirable property of the logarithmic pooling operator is log-concavity.
Log-concavity of the pooled prior may be important to consider in order to guarantee unimodality and certain conditions on tail behaviour -- see~\cite{Bagnoli2005}.
This motivates the following theorem, which is, to the best of our knowledge, a new result:
\begin{theo}
\label{thm:concavity}
\textbf{Log-concavity}. 
If $\mathbf{F}_{\theta}$ is a set of log-concave distributions, then $\pi(\theta\mid \boldsymbol \alpha)$ is also log-concave.
Moreover, logarithmic pooling is the~\underline{only} pooling operator that will \underline{always} produce a log-concave density when all the elements of $\mathbf{F}_{\theta}$  are log-concave.
\end{theo}
\begin{proof}
 See the Appendix.
\end{proof} 
Theorem~\ref{thm:concavity} tells us that logarithmic pooling is the only aggregation method to universally preserve log-concavity, for any configuration of the weights ($\boldsymbol{\alpha}$).
This universality result is important because it holds for any set of log-concave distributions, $\mathbf{F}_{\theta}$.
In contrast, a linear pool of $K + 1$ Gaussian distributions with common mean, $\pi_{\text{linear}}(\theta) = \sum_{i=0}^K \alpha_i f_i(\theta)$ , would produce a log-concave pooled distribution for any $\boldsymbol{\alpha}$, but this would potentially fail if the means were different.

\subsection{Exponential family}
\label{sec:expofamily}

The exponential family of probability distributions finds widespread in the modelling of empirical phenomena.
In this section we give expressions for the entropy and Kullback-Leibler divergence for the pooled distributions. 
These will be useful in applications presented later in the paper.

Suppose we are interested in a random variable $Y$ whose distribution belongs to the exponential family with parameter $\theta$ and probability density function (pdf) given by
\begin{equation}
\label{eq:exponentialfamily}
f(y|\theta) = h(y) e^{\theta y - s(\theta)}.
\end{equation}

Let $\mathbf{F}_{y}$ be a set of densities on $y$ of the form in~(\ref{eq:exponentialfamily}), $f_i(y|\theta_i)$, $ i = 0, 1, \ldots, K$. 
The combined (log-pooled) distribution also belongs to the exponential family:
\begin{equation}
\label{eq:pooldistEF}
\pi(y| \boldsymbol\alpha ) = t(\boldsymbol\alpha) h^\ast (y) e^{\theta^\ast y - s^\ast (\boldsymbol\theta)},
\end{equation}
where $\boldsymbol\theta :=\{\theta_0, \theta_1, \ldots, \theta_K \}$, $h^\ast (y) = \prod_{i = 0}^K h_i(y)^{\alpha_i}$,  $\theta^\ast = \sum_{i = 0}^K \alpha_i \theta_i$ and $s^\ast (\boldsymbol\theta) = \sum_{i = 0}^K \alpha_i s_i(\theta_i)$.

The entropy function of the log-pooled distribution is
\begin{equation}
\label{eq:entropydistEF}
H_\pi(Y; \boldsymbol\alpha) :=  - \mathbb{E}_{\pi}\left[-\log \pi(Y | \boldsymbol\alpha) \right] = -\log t(\boldsymbol\alpha) + s^\ast (\boldsymbol\theta) - \mathbb{E}_\pi[\log h^\ast (Y)] - \theta^\ast \mathbb{E}_\pi[Y] \: ,
\end{equation}
where $\mathbb{E}_{\pi}\left[ g(Y) \right]$ is the expectation of a $\pi$-measurable function $g(Y)$ with respect to $\pi( y | \boldsymbol\alpha)$, when the integral exists.

The Kullback-Leibler divergence between the pooled distribution (\ref{eq:pooldistEF}) and each distribution in $\mathbf{F}_{y}$ can be written as:
\begin{equation}
\label{eq:KLdistEF}
\operatorname{KL}(\pi || f_i )  =  - H_\pi(Y; \boldsymbol\alpha) - \mathbb{E}_\pi[\log h_i(Y)] - \theta_i \mathbb{E}_\pi[Y] + s_i(\theta_i).
\end{equation}

These expressions allow for easy computation of information measures for a broad class of distributions, which will be useful in the remainder of this paper (see also Appendix~\ref{sec:appendix_common_poolings}).

\subsubsection{Conjugate priors to the exponential family}
\label{sec:conjugexpofamily}

A conjugate prior family for $f(y|\theta)$ (\ref{eq:exponentialfamily}), has the following form~\citep{Diaconis1979}:
\begin{equation}
\label{eq:priorEF}
g(\theta | a, b) = K(a,b) e^{\theta a - b s(\theta)} \: ,
\end{equation}
where $K(a,b)$ is a normalising constant.
Similar to the above, let $\mathbf{G}_{\theta}$ be a set of log-conjugate prior distributions representing the opinions of $K+1$ experts, and $g_i(\theta) = g(\theta | a_i, b_i)$ from equation (\ref{eq:priorEF}).

The log-pooled prior is also a conjugate prior for $f(y|\theta)$ with hyperparameters given by a weighted mean of the experts's hyperparameters, i.e., $\pi(\theta|\boldsymbol\alpha) = g(\theta | a^*, b^* )$, where $a^* = \sum_{i=0}^K \alpha_i a_i$ and $b^* = \sum_{i=0}^K \alpha_i b_i$.

The entropy function of the log-pooled prior (\ref{eq:priorEF}) is given by
\begin{equation}
\label{eq:entropypriorEF}
H_\pi(\theta; \boldsymbol\alpha) = - \log (K(a^*, b^*))  -  a^*  \mathbb{E}_\pi[\theta]  +  b^*  \mathbb{E}_\pi[s(\theta)] \: .
\end{equation}

And the Kullback-Leibler divergence, $KL(\pi || g_i )$, is the following
\begin{equation}
\label{eq:KLpriorEF}
KL( \pi || g_i ) = - H_\pi(\theta; \boldsymbol\alpha) - \log( K(a_i,b_i)) - a_i \mathbb{E}_\pi[\theta] + b_i \mathbb{E}_\pi[s(\theta)] \: .
\end{equation}

\subsection{Bayesian melding}
\label{sec:background_melding}

Another important application of logarithmic pooling is in the Bayesian melding method of~\cite{Poole2000}.
Deterministic simulation models are widespread in Science and Engineering (see~\cite{Poole2000} and references therein).
One is often interested in a deterministic model $M$ with inputs $\theta \in \boldsymbol\Theta \subseteq \mathbb{R}^p$ and outputs $\phi \in \boldsymbol\Phi\subseteq \mathbb{R}^q$, such that $\phi = M(\theta)$.
If one wants to learn about $\theta$ from data and a (prior) distribution on $\phi$ is available, then one needs a method to combine the information between the prior on $\theta$ and the prior induced on it  through $M$, which is often non-invertible.

Bayesian melding seeks to draw inference by first employing logarithmic pooling to construct a prior on $\phi$ of the form
\begin{equation}
 \label{eq:BMpoolprior}
 \tilde{q}_{\Phi}(\phi) \propto q_1^\ast(\phi)^\alpha q_2(\phi)^{1-\alpha},
\end{equation}
where $q_1^\ast()$ is the \textbf{induced} prior on the outputs and $q_2$ is the prior on $\phi$ without considering the deterministic model, henceforth called the natural prior on $\phi$.
The prior in~(\ref{eq:BMpoolprior}) can then be inverted to obtain a \textit{coherised} prior on $\theta$, $\tilde{q}_{\Theta}(\theta)$.
\cite{Poole2000} give  a way of obtaining $\tilde{q}_{\Theta}$ even when $M$ is non-invertible, which we will not discuss further here. 

Standard Bayesian inference may then follow,  leading to the posterior
\begin{equation}
 \label{eq:BMpoolposterior}
 p_{\Theta}(\theta) \propto \tilde{q}_{\Theta}(\theta) L_1(\theta) L_2(M(\theta)),
\end{equation}
which enjoys all the properties of usual posterior distributions.
The method allows standard Bayesian inference to be carried out about all quantities of interest in the model, which makes it attractive to application in policy making~\citep{Alkema2008}, where proper acknowledgment of uncertainty is crucial.

In~\cite{Poole2000} (Section 6.2 therein), the authors fix $\alpha = 1/2$, justifying their choice by the fact that while the weights should reflect the reliability of each expert (information source).
In their Bayesian melding analysis, one is combining distributions based on different bodies of evidence, but assessed by the same expert.
Another option is to fix $\alpha = 1-\epsilon$, with $\epsilon$ small~\citep{Alkema2007}.
This can be useful when the prior distribution on outputs is uniform, as it still enforces the constraint, but keeps the prior information about the inputs.
Here we relax the restriction of fixing the weight, instead modelling $\alpha$ through a hyperprior -- see Sections~\ref{sec:bowhead} and~\ref{sec:SIR_flu}. 

\section{Assigning the weights in logarithmic pooling}
\label{sec:weights}

The weights ($\boldsymbol \alpha$) play a key role on the logarithmic pooling and hence their choice is critical.
Building on work by~\cite{Poole2000,Rufo2012A,Rufo2012B} and~\cite{Abbas2009}, we now move on to study three approaches to assigning the weights in logarithmic pooling.
The first two approaches are based on optimality criteria and a third method proposes assigning a (hyper)prior to the weights.

\subsection{Choosing weights based on optimality criteria}

The first set of approaches we will consider attempt to assign the weights by achieving an optimality condition using only information contained in the expert distributions themselves, without reference to any external information such as observed data.

\subsubsection{Maximising entropy}
\label{sec:maxent}

In a context of near complete uncertainty about the relative reliabilities of the experts (information sources) it may be desirable to combine the prior distributions such that $\pi(\theta)$ is maximally diffuse. 
According to its proponents, such an approach would ensure that, given the constraints imposed by $\mathbf{F}_{\theta}$, the pooled distribution is the one which best represents the current state of knowledge~\citep{Jaynes1957,Savchuk1994}.
In order to choose $\boldsymbol\alpha$ so as to maximise prior 
diffuseness, one can maximise the entropy of the log-pooled prior, i.e.
\begin{align}
\nonumber
H_{\pi}(\theta; \boldsymbol\alpha) &= \mathbb{E}_{\pi}\left[-\log \pi(\theta) \right] =-\int_{\boldsymbol\Theta}\pi(\theta)\log\pi(\theta)\, d\theta,\\
\label{eq:entropypiB}
&= -\sum_{i=0}^{K} \alpha_i \mathbb{E}_{\pi}[\log f_i] - \log t(\boldsymbol\alpha).
\end{align}
Formally, we want to find $\hat{\boldsymbol\alpha}$ such that
\begin{equation}
\label{eq:argmaxEnt}
 \hat{\boldsymbol\alpha}:= \argmax_{\boldsymbol\alpha} H_{\pi}(\theta; \boldsymbol\alpha).
\end{equation}

This approach, however, does not result in a convex optimisation problem, therefore one is not guaranteed to find a unique solution -- see Remark~\ref{rmk:uniqueness} for intuition as to why.
A possible resolution to the non-uniqueness of the maximum entropy solution would be to add further constraints, for instance requiring that $E_{\pi}[\theta] = m$.
It is however unclear which set of constraints would ensure uniqueness.

\subsubsection{Minimising Kullback-Leibler divergence}
\label{sec:minKL}

One could also wish to choose the pooling weights so as to minimise the total Kullback-Leibler divergence between the pooled distribution, $\pi$, and each distribution in $\mathbf{F}_{\theta}$.
Let $E_i[g]$ be the expectation of a measurable function $g: \boldsymbol{\Theta} \to \mathbb{R}$ with respect to each density $f_i$.
We can define a loss function such that
\begin{align}
\nonumber
L(\boldsymbol\alpha) &= \sum_{i=0}^K  \text{KL}(f_i || \pi ), \\
\label{eq:KLexpanded}
     &= - (K+1)\log t(\boldsymbol\alpha) - (K+1) \sum_{i=0}^K\alpha_i \mathbb{E}_i [\log f_i ]  + \sum_{i=0}^K \mathbb{E}_i\left[\log f_i\right], 
\end{align}
and we want to find 
\begin{equation}
\label{eq:argminKL}
    \hat{\boldsymbol\alpha}:= \:\argmin_{\boldsymbol\alpha} L(\boldsymbol\alpha).   
\end{equation}
Fortunately, this set up leads to a unique solution, a result we summarise in Remark~\ref{rmk:uniqueness}.
\begin{remark}
\label{rmk:uniqueness}
\textbf{Uniqueness of the minimum KL solution}.
 The distribution obtained following~(\ref{eq:argminKL}) is unique, i.e., there is only one aggregated prior $\pi(\theta \mid \boldsymbol\alpha)$ that minimizes $L(\boldsymbol\alpha)$.
\end{remark}
\begin{proof}
We begin by noting that the second term in~(\ref{eq:KLexpanded}) is a linear combinations of the weights, and hence we may restrict attention only to the first term -- the third term does not depend on $\boldsymbol{\alpha}$.
Next, recall that minimising~(\ref{eq:KLexpanded}) is equivalent to maximising $\log t(\boldsymbol\alpha) = \log\int_{\boldsymbol\Theta}\prod_{i=0}^{K}f_i(\theta)^{\alpha_i}\, d\theta$.
Proposition 3.1 in~\cite{Rufo2012A} states that $t(\boldsymbol\alpha)$ is (log-)concave, therefore any optimisation problem which involves minimising $-\log t(\boldsymbol{\alpha})$ is convex.
We thus conclude that the problem in~(\ref{eq:argminKL}) has a unique solution.
\end{proof}
By contrast, the problem in~(\ref{eq:argmaxEnt}) requires one to minimise $\ln t(\boldsymbol\alpha)$, hence lacking a sufficient condition for the existence of a unique solution.
Likewise, using  the loss function $L^\prime(\boldsymbol\alpha) = \sum_{i=0}^K  \text{KL}(\pi ||f_i )$ would not lead to a unique solution.
See Appendix~\ref{sec:appendix_compdetails} for implementation details.

\subsection{Hierarchical modelling of the weights}
\label{sec:hierPrior}

As discussed by~\cite{Poole2000} and others~\citep{Zhong2015,Li2017}, estimating the weights would be of interest since this would allow one to assess the reliability of each source of information (expert).
\cite{Li2017} explore the idea of computing the pooled distribution for several values of the weights.
Whilst informative, this approach has two issues: (a) it does not scale well with increasing the number of distributions being combined, $K$, and; (b) it fails to account for any (posterior) dependence between model parameters and the weights.
In this section we propose placing a hierarchical prior on the weights, allowing for standard Bayesian inference about these quantities.

A natural choice for a prior distribution for $\boldsymbol\alpha$ is the $(K+1)-$dimensional Dirichlet distribution
\begin{equation}
 \label{eq:generalcondprior}
 \pi_A(\boldsymbol\alpha) = \frac{1}{\mathcal{B}(\boldsymbol x)}\prod_{i=0}^K \alpha_i^{x_i-1},
\end{equation}
where $\boldsymbol x = \{ x_0, x_1, \ldots, x_K\}$ is the vector of hyperparameters for the Dirichlet prior and $\mathcal{B}(X)$ is the multinomial beta function.
The Dirichlet offers a simple, albeit potentially inflexible prior.

A more flexible prior for $\boldsymbol\alpha$ is the logistic-normal distribution~\citep{Aitchson1980}:
\begin{equation}
 \label{eq:aitchinsonprior}
 \pi_A(\boldsymbol\alpha \mid \boldsymbol \mu, \boldsymbol \Sigma) = \frac{1}{|2\pi \boldsymbol \Sigma|^{\frac{1}{2}}}\frac{1}{\prod_{i=0}^K \alpha_i}
  \exp\left(
     \left(\log\left(\frac{\boldsymbol \alpha_{-K}}{\alpha_K}\right) - \boldsymbol \mu\right)^T
     {\boldsymbol \Sigma}^{-1}
     \left(\log\left(\frac{\boldsymbol \alpha_{-K}}{\alpha_K}\right) - \boldsymbol \mu\right)
     \right),
\end{equation}
where $\boldsymbol \alpha_{-K}$ represents the vector $\boldsymbol \alpha$ without the $K$-th element, $\boldsymbol \mu$ is a $K$-size mean vector, and $\boldsymbol \Sigma$ is a $K \times K$ covariance matrix.
\citep{Aitchson1980} propose choosing $\boldsymbol \mu$ and $\boldsymbol \Sigma$ minimizing the KL divergence between the Dirichlet (\ref{eq:generalcondprior}) and the logistic-normal (\ref{eq:aitchinsonprior}) distributions, i.e.
\begin{align}
 \label{eq:momentmatching}
 \mu_i & = \psi(x_i) - \psi(x_K), \quad i=0,1,\ldots,K-1, \\
 \Sigma_{ii} & = \psi'(x_i) + \psi'(x_K), \quad i=0,1,\ldots,K-1, \\
 \Sigma_{ij} & = \psi'(x_K),
\end{align}
where $\psi(\cdot)$ is the digamma function, and $\psi'(\cdot)$ is the trigamma function.

The marginal prior for $\theta$,
\begin{equation}
\label{eq:marginalbeta}
\tilde{\pi}(\theta) = \int_{\mathcal{A}} \pi(\theta \mid \boldsymbol\alpha) \pi_A(\boldsymbol\alpha)d\boldsymbol\alpha,
\end{equation}
can also be efficiently approximated through Monte Carlo sampling when $\pi$ can be written in closed-form.
Even when it cannot be expressed analytically, it is still possible to sample from the marginal prior by using quadrature-based methods for computing $t(\boldsymbol\alpha)$ when $\theta$ is unidimensional (see Discussion).

Concerning posterior inference, the marginal posterior for $\theta$ can be obtained through standard methods and shall not be discussed further.
The next object to consider is the marginal posterior for the weights, $p(\boldsymbol\alpha \mid \boldsymbol{x})$, which can be obtained through
\begin{align}
\nonumber
 p(\boldsymbol\alpha \mid \boldsymbol{x}) &= \int_{\boldsymbol\Theta} p(\boldsymbol\alpha, \theta \mid \boldsymbol{x})\,d\theta, \\
 \nonumber
 &= \int_{\boldsymbol\Theta} \frac{L(\boldsymbol{x} \mid \theta) \pi(\theta \mid \boldsymbol\alpha)\pi_A(\boldsymbol\alpha)}{c(\boldsymbol{x})}\,d\theta,\\
 \nonumber
 &= \frac{\pi_A(\boldsymbol\alpha)}{c(\boldsymbol{x})} \int_{\boldsymbol\Theta} L(\boldsymbol{x} \mid \theta) \pi(\theta \mid \boldsymbol\alpha)\,d\theta,\\
 \label{eq:marginal_posterior_alpha}
 &\propto \pi_A(\boldsymbol\alpha) \kappa(\boldsymbol\alpha, \boldsymbol{x}),
\end{align}
where $c(\boldsymbol{x}) := \int_{\mathcal{A}}\int_{\boldsymbol\Theta} p(\boldsymbol\alpha, \theta \mid \boldsymbol{x})\,d\theta\,d\boldsymbol{\alpha}$.

In some situations, in particular the conjugate situation discussed in Section~\ref{sec:conjugexpofamily} and exemplified in the Applications section below, it is possible to write down $\kappa(\boldsymbol\alpha, \boldsymbol{x})$ in closed-form.
This is very convenient because the posterior expectation of the weights, $E_p[\boldsymbol\alpha \mid \boldsymbol{x}]$, becomes $E_{\pi_A}[\boldsymbol\alpha \kappa(\boldsymbol\alpha, \boldsymbol{x})]$, i.e., the expectation of a known function with respect to the prior on the weights.
This expectation can be easily and accurately approximated with simple Monte Carlo techniques rather than MCMC -- see Section~\ref{sec:learning_rate} for example applications.

\section{Applications}
\label{sec:apps}

In this section we shall present a wide range of applications for logarithmic pooling, from prior elicitation to meta-analysis to Bayesian melding.
Computational details, along with instructions to get reproducible code, are given in  Appendix~\ref{sec:appendix_compdetails}.

\subsection{Elicitation: combining expert priors on survival probabilities}
\label{sec:survivalProbs}

The first example we consider is combining expert opinions about probabilities and proportions.
We analyse an example proposed by~\cite{Savchuk1994} (also discussed in~\cite{Rufo2012B}) in which four experts are required supply prior information about the survival probability $\theta$ of a certain unit.
The experts express their opinion as prior means for the survival probability, which~\cite{Savchuk1994} then use to construct prior distributions with maximum variance given the restriction on the means.
From the vector of prior means $\mathbf{m} = \{ m_0 = 0.95, m_1 = 0.80, m_2 = 0.90, m_3 = 0.70 \}$, the authors obtain the parameters of the Beta distributions for each expert,  $\mathbf{a} = \{ a_0 = 18.10, a_1 = 3.44 , a_2 = 8.32, a_3 = 1.98 \}$ and  $\mathbf{b} = \{ b_0 = 0.955 , b_1 = 0.860, b_2 = 0.924, b_3 = 0.848\}$.
Furthermore, an experiment is conducted and $y = 9$ successes out of $n = 10$ trials are observed.
Thus, in this application we are able to estimate the posterior distribution for the survival probability and also, with  the hierarchical modelling approach, the posterior distribution for the weights in face of the observed data.
For the hierarchical priors, we employ a $\text{Dirichlet}(1/10, 1/10, 1/10, 1/10)$ and a moment-matching logistic-normal priors (see Section~\ref{sec:learning_rate} for justification).

The probability distribution of the survival probability for the $i$-th expert is a Beta distribution with (hyper)parameters $a_i$ and $b_i$.
The log-pooled distribution for $\theta$ is then
\begin{align}
\nonumber
\pi(\theta) & \propto \prod_{i=0}^{K}f_i(\theta;a_i,b_i)^{\alpha_i},\\
\nonumber
            & \propto \prod_{i=0}^{K} \left(\theta^{a_i-1}(1-\theta)^{b_i-1} \right)^{\alpha_i},\\
\label{eq:betabern}
&\propto \theta^{a^*-1}(1-\theta)^{b^*-1},
\end{align}
with $a^* =\sum_{i=0}^{K}\alpha_ia_i$ and $b^* = \sum_{i=0}^{K}\alpha_ib_i$.
Note that (\ref{eq:betabern}) is the kernel of a Beta distribution with parameters $a^*$ and $b^*$. Hence the entropy is the following
\begin{equation}
 \label{eq:entropybeta}
 H_{\pi}(\theta) = \log \mathcal{B}(a^*,b^*) - (a^*-1)\psi(a^*) - (b^*-1)\psi(b^*) + (a^*+b^* -2)\psi(a^*+b^*).
\end{equation}
And the KL divergence between $\pi(\theta)$ and $f_i(\theta)$  is
\begin{equation}
\begin{split}
 \label{eq:KLbeta}
 d_i = KL(f_i || \pi) = \ln\left(\frac{\mathcal{B}(a^*, b^*)}{\mathcal{B}(a_i, 
b_i)}\right) & + (a_i - a^*) \psi(a_i)+ (b_i - b^*)\psi(b_i) \\
 &- (a_i-a^* + b_i - b^*)\psi(a_i + b_i).
\end{split}
\end{equation}
In this conjugate setting, the posteriors associated with each expert are also Beta distributions with
parameters $a_i^\prime = a_i + y$ and $b_i^\prime = b_i + (n-y)$.
This allows us to employ the maximum entropy and minimum KL procedures to combine these posterior distributions and thus make the weights comparable with the posterior means obtained with the hierarchical priors.

Our analysis of this example is thus split into two: weights for the priors and for the posteriors.
Before observing any data, we can employ the optimisation procedures discussed above to obtain weights only taking into account information encoded in the expert priors themselves.
To these optimisation procedures we add the technique of~\cite{Rufo2012B} which seeks to minimise KL distance between the pooled prior and the Jeffreys's posterior.
When data are available, we can then use maximum entropy and minimum KL to obtain the weights in the same fashion as before, but now  also estimate the posterior distribution of weights using a hierarchical prior.
Finally, for this example we can also compute the integrated (marginal) likelihood of each expert, meaning that we can, assuming one of the experts is correct, compute ``model'' probabilities by normalising the marginal likelihoods (see Section~\ref{sec:learning_rate}, below).

In Table~\ref{tab:alphasBeta} we present weights obtained with the optimisation methods for the priors, including the solution found by~\cite{Rufo2012B} (Section 5.2 therein).
With regard to the posteriors, we show maximum entropy, minimum KL along with posterior means of the weights under two prior distributions (Dirichlet and logistic-normal).
Maximising the entropy of the pooled prior -- and posterior -- lead to the degenerate solution $\boldsymbol \alpha = \{0, 0, 0, 1 \}$, which gives all the weight to the most diffuse prior distribution -- $\text{Beta}(1.98, 0.848)$.
Since $t(\boldsymbol\alpha)$ is concave, we expect to find the maximum entropy given by the boundary conditions, which may lead to points in the border of the simplex.
Unsurprisingly, the same solution was found by~\cite{Rufo2012B}, whose method tends to favour more diffuse distributions.
Minimising Kullback-Leibler divergence between the pooled prior and each expert prior leads to a unique solution but in this case also suggests to discard two of the opinions.

The hierarchical priors gave very similar posterior distributions for the weights, which assign the experts nearly equal weight, although the logistic-normal prior lead to results closer to the marginal likelihood-based weights.

\begin{table}[ht]
\caption{\textbf{Weights obtained using different methods for the survival probability example~\citep{Savchuk1994}.}
$^1$ -- Kullback-Leibler; $^2$ -- Posterior mean for $\boldsymbol\alpha$.}
\centering
\begin{tabular}{cccccc}
\hline
                           & Method                & $\alpha_0$   & $\alpha_1$   & $\alpha_2$   & $\alpha_3$\\
\hline
\multirow{3}{*}{Prior}     & Maximum entropy       & 0.00 & 0.00 & 0.00 & 1.00  \\
                           & Minimum KL$^1$        & 0.04 & 0.96 & 0.00 & 0.00 \\
                           & \cite{Rufo2012A}      & 0.00 & 0.00 & 0.00 & 1.00  \\
                           &                       &      &      &      &      \\
\multirow{5}{*}{Posterior} & Maximum entropy       & 0.00 & 0.00 & 0.00 & 1.00  \\
                           & Minimum KL            & 0.17 & 0.83 & 0.00 & 0.00 \\
                           & Dirichlet$^2$        & 0.26 & 0.24 & 0.27 & 0.23 \\
                           & Logistic-normal$^2$ & 0.27 & 0.24 & 0.31 & 0.18 \\
                           & Marginal likelihoods  & 0.27 & 0.24 & 0.30 & 0.19\\
\hline                
\end{tabular}
\label{tab:alphasBeta}
\end{table}

\begin{figure}[!ht]
\begin{center}
\subfigure[][Expert priors]{\includegraphics[scale=.45]{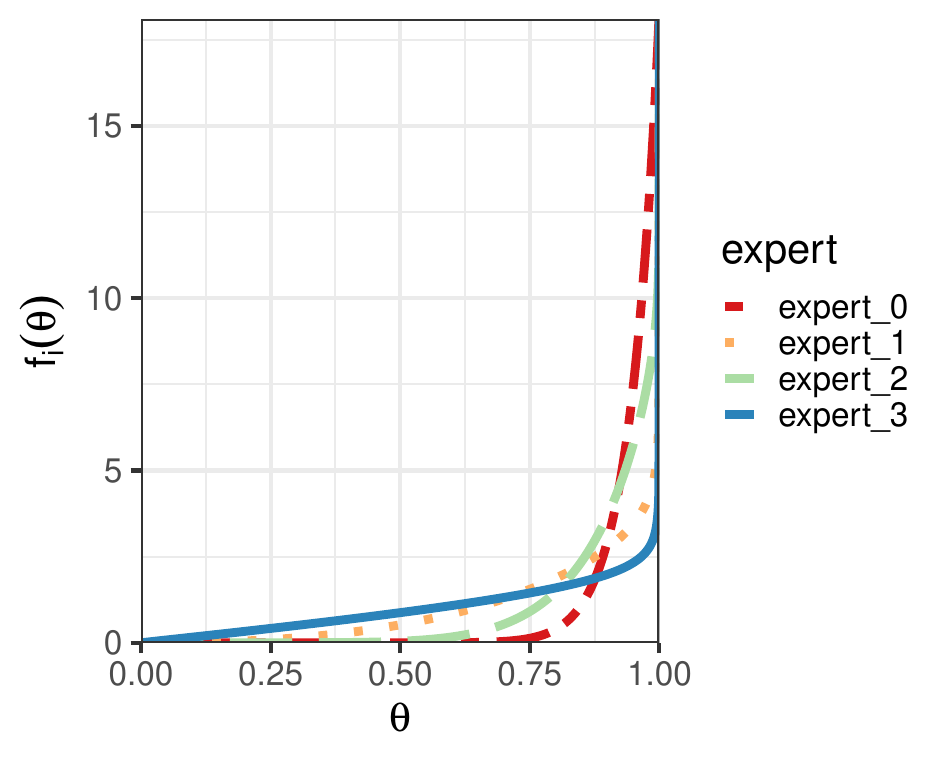}}
\subfigure[][Pooled priors]{\includegraphics[scale=.45]{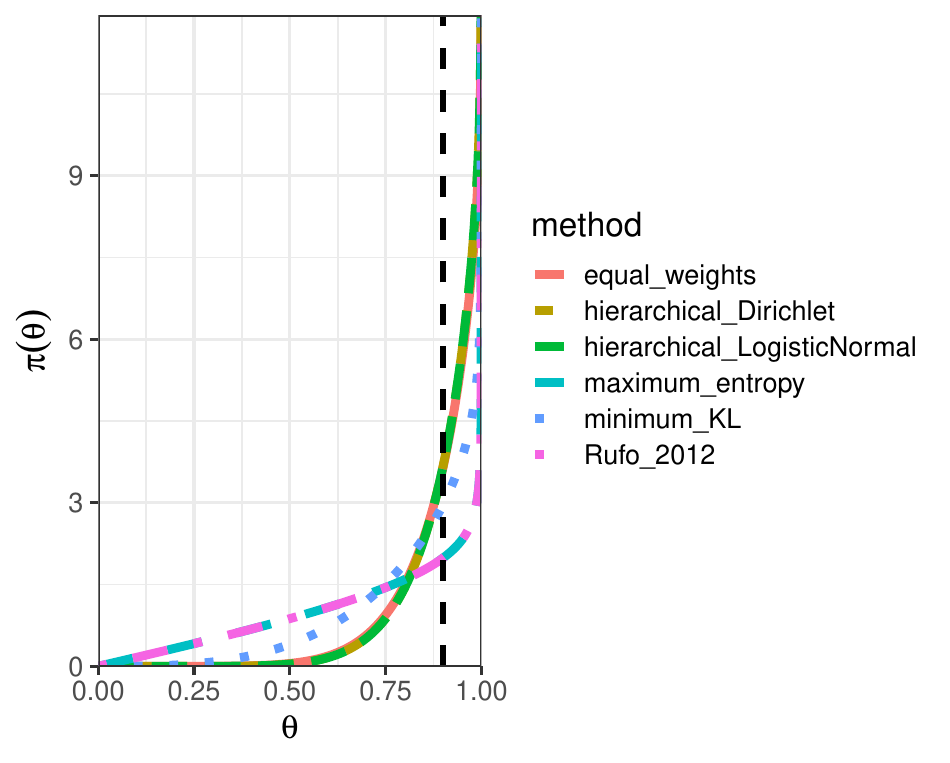}}\\
\subfigure[][Expert posteriors]{\includegraphics[scale=.45]{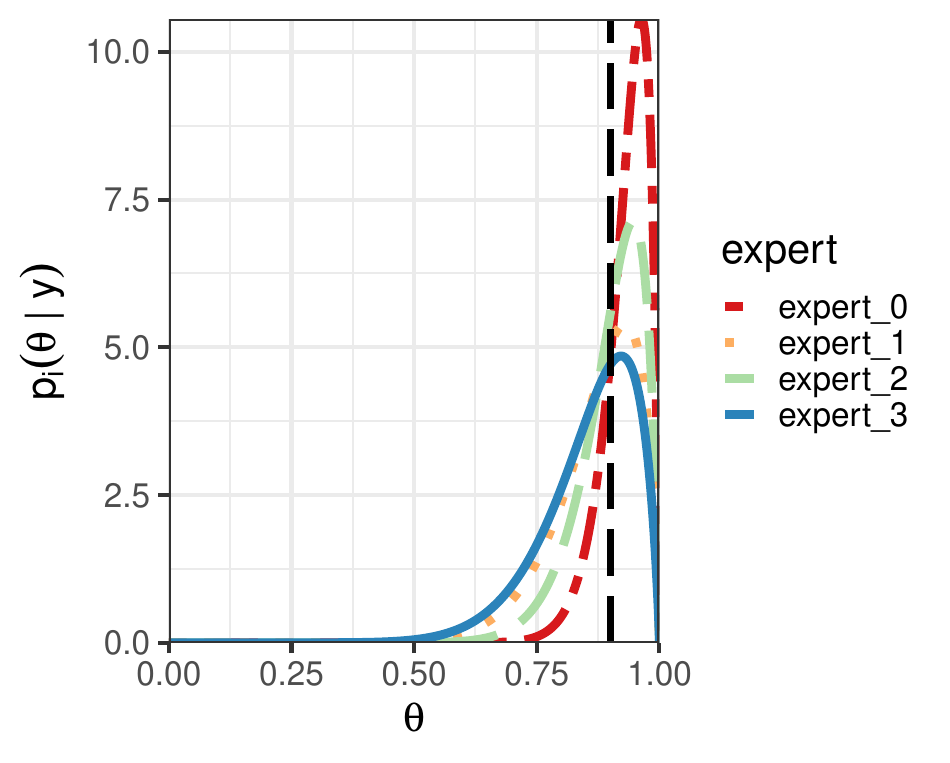}}
\subfigure[][Pooled posteriors]{\includegraphics[scale=.45]{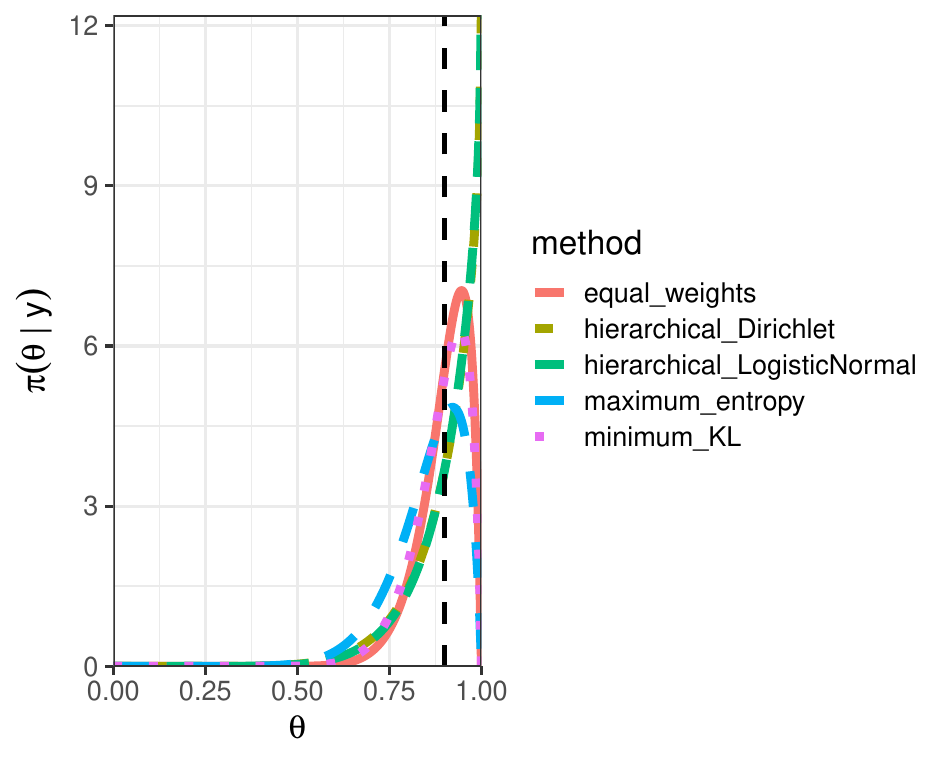}}
\end{center}
\caption{\textbf{Prior and posterior densities for the survival probability $\theta$}.
Panel (a) shows the distributions elicited by each expert (data from~\cite{Savchuk1994}) and panel (b) shows the pooled priors and posteriors obtained using the methods discussed in this paper and the solution found by~\cite{Rufo2012B}.
Panels (c) and (d) show the corresponding plots for the posterior distributions after observing the data, $y = 9, n = 10$.
The black dashed vertical line marks the maximum likelihood estimate $\hat{\theta}= 9/10$.
}
\label{fig:priors_pooled_Savchuk}
\end{figure}

Figure~\ref{fig:priors_pooled_Savchuk} shows the prior densities for each expert and pooling method and Table~\ref{tab:prior_posteriorsSavchuk} contains the prior and posterior mean and credibility intervals from each of the methods and also the case in which we assign an equal weight ($1/K$) to each opinion.
Assigning equal weights actually gives a prior mean that is the same as the maximum likelihood estimate of $\theta$, $\hat{\theta} = 9/ 10$.
This explains why both hierarchical posteriors resemble equal weights so closely.
Finally, we use the integrated (marginal) likelihood (\cite{Raftery2007}, eq. 9), $l(y) = \int_{0}^{1}f(y|\theta)\pi(\theta)\, d\theta$, as a univariate summary to compare the priors.
The marginal likelihood for the $i$-th expert and $J$ observations of the form $\{ y_j, n_j\}$ is:
\begin{align}
  \label{eq:marglike}
l_i(y_j, n_j) &= \int_{0}^{1}\mathcal{L}(\theta|y_j, n_j)\pi_i(\theta)\, d\theta\nonumber\\
 &= \prod_{j = 1}^{J}\frac{\Gamma(n_j-1)}{\Gamma(n_j-y_j + 1)\Gamma(y_j+1)}\frac{\Gamma(a_i + b_i)}{\Gamma(a_i + b_i + n_j)}\frac{\Gamma(a_i + y_j)}{\Gamma(a_i)}\frac{\Gamma(b_i + n_j - y_j) }{\Gamma(b_i)}.
 \end{align}
For the hierarchical priors we take the posterior mean of $(a^\star, b^\star)$ as $(a_i, b_i)$.
Results are given in Table~\ref{tab:marglikes} and show that, apart from expert $3$ -- and hence the maximum entropy pooled prior--, all other pooled priors and individual experts' priors give similar marginal likelihoods.

The posterior distribution for the weights estimated under both priors favours expert 2, the expert with the highest marginal likelihood.
The logistic-normal gives expert 2 a higher weight when compared with the Dirichlet.
This is connected to the increased flexibility of the logistic-normal (see Section~\ref{sec:learning_rate}).

We stress that that the marginal likelihoods are not being used here as a means of selecting priors, but rather as a useful univariate summary which is informative about the compatibility with the observed data and hence informative about prior-data conflict.
While in this example one can gain insight into prior-data conflict from just the prior means and $y/n$, in other situations it might be harder to discern which expert gave the best (prior) guess.

\begin{table}[ht]
\caption{\textbf{Prior and posterior mean and credibility intervals for each method for assigning the weight, survival probability example~\citep{Savchuk1994}.}
Values for the hierarchical priors are from the marginal prior of $\theta$ in~(\ref{eq:marginalbeta}).
}
\centering
\begin{tabular}{ccc}
 \hline
Method & Prior & Posterior  \\ 
 \hline
 Equal weights & 0.90 (0.64--1.00) & 0.90 (0.73--0.99) \\ 
 Maximum entropy &  0.70 (0.17--0.99) & 0.86 (0.63--0.98) \\ 
 Minimum KL  &  0.82 (0.42--1.00) & 0.87 (0.67--0.99) \\ 
 \cite{Rufo2012B} & 0.70 (0.17--0.99) & 0.86 (0.63--0.98)\\
 Dirichlet  & 0.86 (0.40--1.00) & 0.89 (0.70--0.99) \\ 
 Logistic-normal & 0.88 (0.35--1.00) & 0.89 (0.71--0.99) \\ 
  \hline
\end{tabular}
\label{tab:prior_posteriorsSavchuk}
\end{table}

\begin{table}[ht]
\caption{\textbf{Integrated likelihoods for the priors of each expert as well as the combined priors, failure probability example}.
For the hierarchical priors we take the posterior expectations of $a^\star$ and $b^\star$ as $a_i$ and $b_i$, respectively.
$^1$ Calculated using the posterior mean of $\boldsymbol\alpha$. }
\centering
\begin{tabular}{cccc}
   \hline
   \multicolumn{2}{c}{Expert priors} &  \multicolumn{2}{c}{Pooled priors} \\
   \hline
   Expert 0 & 0.237 & Equal weights & 0.254\\
   Expert 1 & 0.211 & Maximum entropy & 0.163 \\
   Expert 2 & 0.256 & Minimum KL & 0.223 \\ 
   Expert 3 & 0.163 & Hierarchical prior$^1$ (Dirichlet/logistic-normal) & 0.255 \\
   \hline
\end{tabular}
\label{tab:marglikes}
\end{table}

\subsection{Meta-analysis: HIV prevalence among MSM populations in Brazil}
\label{sec:metaAnalysis}

Another potential application of logarithmic pooling is in meta analysis.
Logarithmic pooling can also be used to combine probability distributions of a particular outcome estimated from several studies. 
In epidemiology, systematic review and meta analysis are popular tools for merging and contrasting results across multiple studies~\citep[Chapter 33]{Rothman2008}.
Moreover, estimation of disease prevalence and the effect of exposure variables are amongst the most important application of meta-analyses in epidemiology.
We illustrate the different approaches to assign weights in the logarithmic polling in the systematic review and meta analysis conducted by~\citet{Malta2010}. 
They analysed studies published from 1999 to 2009 assessing the HIV prevalence among men who have sex with another men (MSM) in Brazil. 
The authors have found six studies that estimated HIV prevalence in MSM population in Brazil. 
Data from each study consists of $n_i$ observed individuals, $y_i$ of which were infected with HIV.

Assuming a uniform prior for the HIV prevalence among MSM, denoted by $\varphi$, and a binomial model for each study, i.e. $Y_i \sim \text{Binomial}(n_i, \varphi)$. 
The posterior distribution for the HIV prevalence conditional on each study is then a Beta distribution with parameters $a_i = y_i + 1$ and $b_i = n_i - y_i + 1$, for $i=0,1, \ldots, 5$.
For the first part of our analysis of this problem we will assume $\boldsymbol F^{B}_\varphi$ to be composed of these posterior distributions.
In meta-analysis it is common for researchers to employ a Gaussian (normal) distribution instead of a distribution with support on $(0, 1)$, relying on the large sample normal approximation of the binomial distribution. 
Here we study how this choice of representation impacts the logarithmic pooling procedure by comparing the Beta and Gaussian distributions as representations of the HIV prevalence among MSM.

If the  probability density on $\varphi$ for each study is now
$$ f_i(\varphi; m_i, v_i) = \frac{1}{\sqrt{2\pi v_i}} \exp\left(\frac{-(\varphi-m_i)^2}{2v_i}\right), $$
where $m_i = y_i/n_i$ and $v_i = m_i(1-m_i)/n_i$.
We have
\begin{align}
\nonumber
\pi(\varphi \mid \boldsymbol\alpha)&= t(\boldsymbol\alpha)\prod_{i=0}^{K}f_i(\varphi; m_i, v_i)^{\alpha_i},\\
\nonumber
&\propto \prod_{i=0}^{K} \left[ \exp\left(\frac{-(\varphi-m_i)^2}{2v_i}\right) \right]^{\alpha_i},\\
&\propto \exp\left[-\frac{1}{2}\left\{\varphi\sum_{i=0}^K\frac{\alpha_i}{v_i} - 2\varphi\sum_{i=0}^K \frac{\alpha_im_i}{v_i} - \sum_{i=0}^K\frac{\alpha_im_i^2}{v_i} \right\}\right].
\end{align}
Completing the square shows $\pi(\varphi)$ is the density of a normal distribution with parameters and $m^* = \frac{\sum_{i=0}^K w_im_i}{\sum_{i=0}^K w_i}$ and $v^* = [\sum_{i=0}^K w_i]^{-1}$,  where $w_i = \alpha_i/v_i$.
The entropy function is then:
\begin{equation}
 \label{eq:normalpoolentropy}
 H_{\pi}(\varphi) = \frac{1}{2}\left[ \ln(2\pi e) - \ln\sum_{i=0}^K w_i\right],
\end{equation}
which achieves its maximum when $\alpha_j = 1$ for $v_j = max(v_1, v_2, \ldots, v_K)$ and thus maximising entropy always leads to degenerate solutions.
The Kullback-Leibler divergence between the pooled distribution $\pi(\varphi)$ and each $f_i(\varphi)$ is then
\begin{equation}
 \label{eq:KL_Gaussian}
  \text{KL}( f_i || \pi) = \frac{1}{2}\log\left(\frac{v^\star}{v_i}\right) + \frac{v_i + (m_i-m^\star)^2}{2v^\star} - \frac{1}{2}.
\end{equation}
For the second part of our analysis of this example, we will assume that set of distributions to be combined, $\boldsymbol F^{G}_\varphi$, is composed by the Gaussian distributions described above.

Table \ref{tab:HIV_MSM} contains the sample size for each study, the total of HIV positive observed, and the estimated prevalence using the Beta distribution described above and a Gaussian distribution (see below). 
Note that the estimated prevalences among MSM are very high when compared with the HIV prevalence in the general population, 0.6\% \citep{Malta2010}.
In addition, there is considerable heterogeneity between studies, with (mean) estimates ranging from 6\%~\citep{Tun2008} to 24\%\citep{Sutmoller2002,Barcellos2003}.
\begin{table}[ht]
\caption{\textbf{Data extracted from the systematic review and meta analysis conducted by \citet{Malta2010} assessing the HIV prevalence among MSM in Brazil.}
$n_i$ is the sample size, $y_i$ is the total of HIV-positive participants in the $i$-th study.
Prevalence estimates are presented as mean and 95\% credibility intervals, either from a Beta distribution with parameters $a_i = y_i + 1$ and $b_i = n_i - y_i  +1$ or a Gaussian distribution with $m_i = y_i/n_i$ and $v_i = m_i(1-m_i)/n_i$ (see text).
}
\label{tab:HIV_MSM}
\centering
\begin{tabular}{rlrrcc}
  \hline
  & & & & \multicolumn{2}{c}{Estimated prevalence, $\varphi$ (95\% CI)}
\\
Study & Reference & $n$ & $y$ &  Beta & Gaussian \\ 
  \hline
0 & \cite{Tun2008}            &  658 &  44 & 0.068 (0.050--0.089) & 0.067 (0.048--0.086)\\ 
1 & \cite{Barcellos2003}  &  461 & 111 & 0.242 (0.204--0.282) & 0.241 (0.202--0.280)\\ 
2 & \cite{Carneiro2003} &  621 &  61 & 0.100 (0.077--0.124) & 0.098 (0.075--0.122)\\ 
3 & \cite{Sutmoller2002}       & 1165 & 281 & 0.242 (0.218--0.267) & 0.241 (0.217--0.266)\\ 
4 & \cite{BMH2000}                  &  642 &  57 & 0.090 (0.069--0.113) & 0.089 (0.067--0.111)\\ 
5 & \cite{Harrison1999}     &  849 &  99 & 0.118 (0.097--0.140) & 0.117 (0.095--0.138)\\ 
   \hline
\end{tabular}
\end{table}

In a meta-analytic context it also makes sense to consider giving weights to each study proportional to sample size, with larger studies receiving larger weights and hence more credence.
We have included this weighting, with $\alpha_i = n_i/ \sum_{i =0}^K n_i$, in our analysis, along with the maximum entropy and minimum KL solutions.
The weights obtained by maximising entropy and minimising KL divergence for both the Beta and Gaussian representations of the prevalence information are given in Table~\ref{tab:weights_MSM}.
While the maximum entropy method lead to the same degenerate solution for both distributions, giving all the weight to the study by~\cite{Barcellos2003}, minimising KL divergence lead to the studies by~\cite{Tun2008} and~\cite{Barcellos2003} being given non-zero weights.
Interestingly, while the minimum KL method for the Beta distribution representation lead to roughly equal weights, with the study by~\cite{Tun2008} being slightly favoured, the solution for the Gaussian representation assigned a much larger weight to the distribution from~\cite{Barcellos2003} (see Discussion).

\begin{table}[!ht]
\caption{\textbf{Weights obtained using different methods for the HIV prevalence example}.
Sample size pertains to assigning the  weights based on the normalised sample sizes ($\alpha_i = n_i/ \sum_{i =0}^K n_i$).
}
\centering
\begin{tabular}{cccccccc}
\hline
Method                            &    & $\alpha_0$ & $\alpha_1$ & $\alpha_2$ & $\alpha_3$ & $\alpha_4$ & $\alpha_5$ \\
\hline
\multirow{2}{*}{Maximum entropy} & Beta     & 0 & 1 & 0 & 0 & 0 & 0\\
                                  & Gaussian & 0 & 1 & 0 & 0 & 0 & 0\\
\multirow{2}{*}{Minimum KL divergence}      & Beta & 0.53 & 0.47 & 0 & 0 & 0 & 0 \\
                                  & Gaussian & 0.17 & 0.83 & 0 & 0 & 0 & 0\\
Sample size                      & & 0.15 & 0.11 & 0.14 & 0.27 & 0.15 & 0.19\\
\hline
\end{tabular}
\label{tab:weights_MSM}
\end{table}

\begin{figure}[!ht]
\begin{center}
\subfigure[][Study distributions]{\includegraphics[scale=.45]{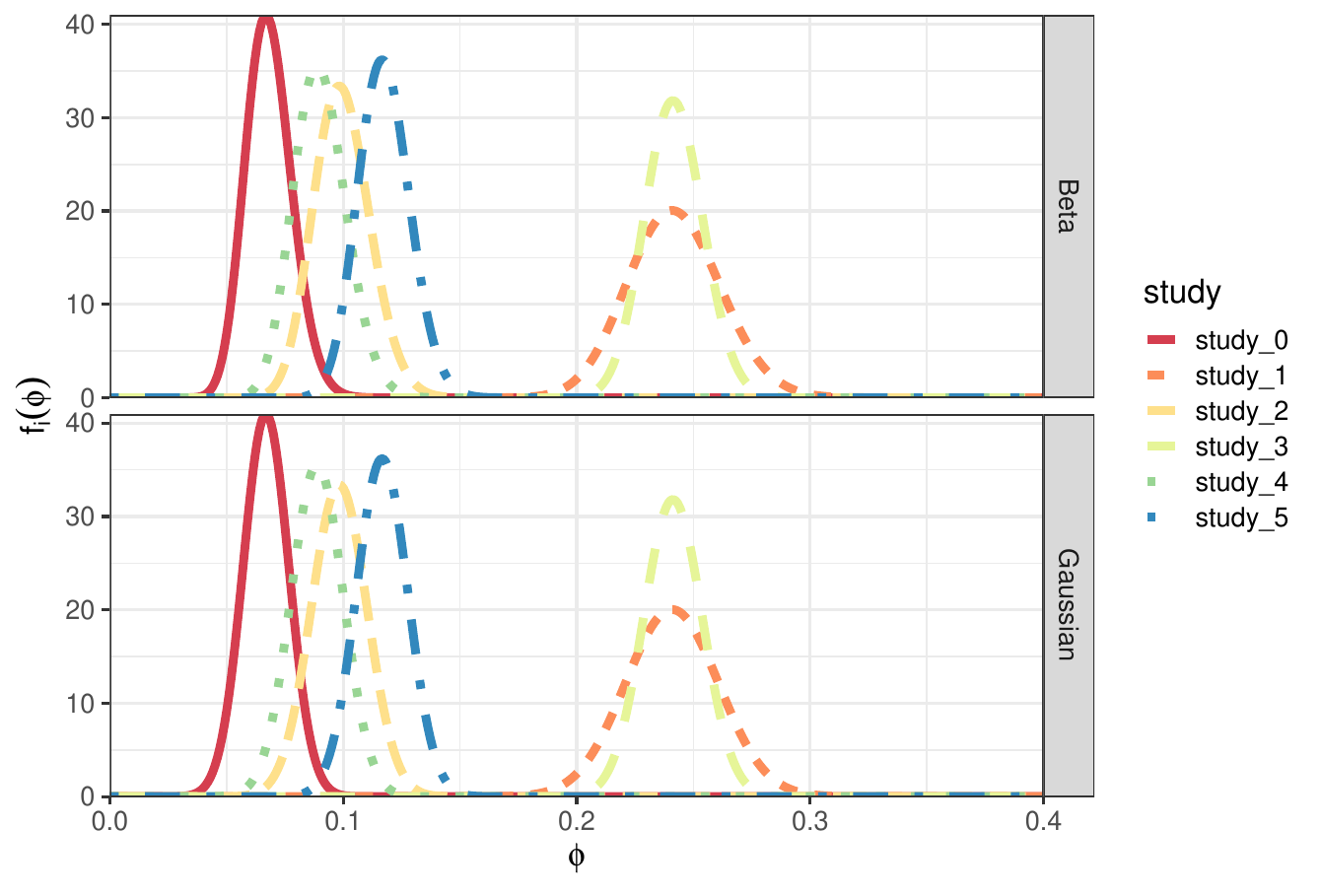}}
\subfigure[][Pooled distributions]{\includegraphics[scale=.45]{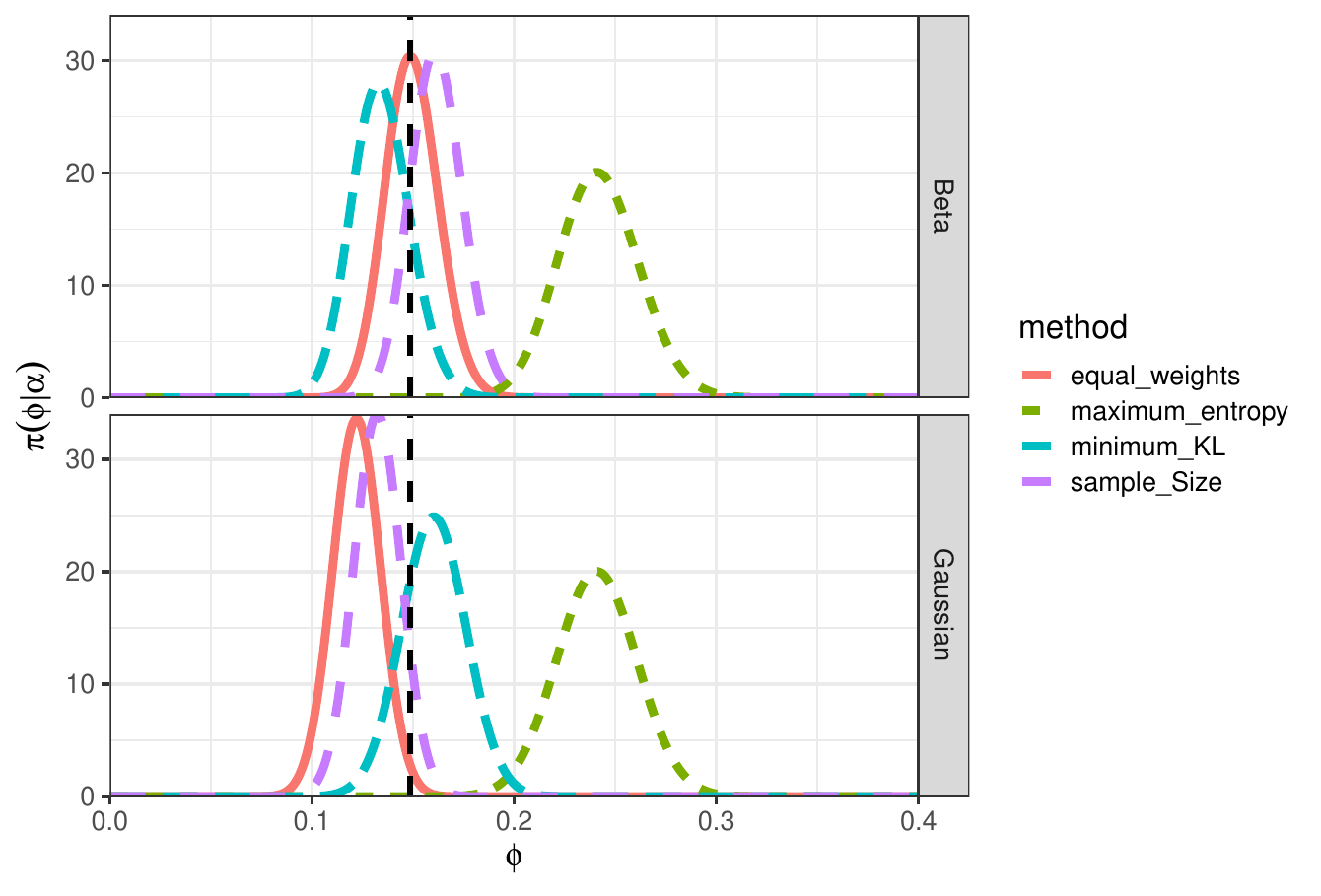}}
\end{center}
\caption{\textbf{Densities for the HIV prevalence among MSM $\varphi$ using Beta and Gaussian distributions}.
Panel (a) shows the distributions obtained from  each study (see Table~\ref{tab:HIV_MSM}) and panel (b) shows the pooled distributions obtained using the methods discussed in this paper.
Vertical tiles show the distribution of choice (Beta or Gaussian) and the vertical dashed black line in panel (b) shows the estimate obtained by combining all studies, $\hat{\varphi} = \sum_{i = 0}^K y_i /\sum_{i = 0}^K n_i$.
}
\label{fig:priors_pooled_MSM}
\end{figure}

Table~\ref{tab:prior_MSM} shows the estimates of the HIV prevalence among MSM in Brazil using different methods for obtaining the pooled distributions and Figure~\ref{fig:priors_pooled_MSM} shows the resulting densities.
For comparison, we also included results from the log-pooled distributions obtained with equal weights and the marginal prior on $\varphi$ induced by the Dirichlet (1/10, 1/10, 1/10, 1/10, 1/10, 1/10) and moment-matching logistic-normal priors on $\boldsymbol\alpha$. 
The heterogeneity observed across studies (Table~\ref{tab:HIV_MSM}) is also reflected in the variation in the combined (pooled) priors under different methods.
In contrast to the mean prevalence of $24\%$ yielded by the maximum entropy pooled prior, all other combined distributions have estimated mean prevalences in the range $[13\%, 16\%]$ for the Beta representation and $[12\%, 16\%]$ for the Gaussian representation.
As expected, the marginal priors for $\varphi$ induced by placing a prior on $\boldsymbol\alpha$ and then marginalising (\textit{via} Monte Carlo) yield broader distributions, which encompass the range of all original studies.
This effect is slightly more pronounced for the logistic-normal prior, pointing towards more flexibility compared to the Dirichlet.

\begin{table}[ht]
\caption{\textbf{Mean and credibility intervals for each method for assigning the weights under two representations of information, HIV prevalence example.}
Sample size pertains to assigning the  weights based on the normalised sample sizes ($\alpha_i = n_i/ \sum_{i =0}^K n_i$).
}
\centering
\begin{tabular}{ccc}
 \hline
 & \multicolumn{2}{c}{Estimated HIV prevalence} \\
Method & Beta & Gaussian  \\ 
 \hline
 Equal weights &  0.150 (0.125--0.176)&  0.122 (0.099--0.145)\\ 
 Maximum entropy & 0.242 (0.204--0.282)  &  0.241 (0.202--0.280)\\ 
 Minimum KL  & 0.134 (0.107--0.163)  & 0.160 (0.129--0.192) \\ 
 Sample size & 0.162 (0.137--0.188) & 0.132 (0.109--0.155)\\
 Dirichlet prior & 0.144 (0.066--0.253) &  0.133 (0.063--0.250)\\ 
 Logistic-normal prior  & 0.143 (0.060--0.261) & 0.138 (0.059--0.259)\\ 
  \hline
\end{tabular}
\label{tab:prior_MSM}
\end{table}

\subsection{Posterior distribution of the weights: interpretability and prior sensitivity}
\label{sec:learning_rate}

The analysis of the survival probabilities in Section~\ref{sec:survivalProbs} instigates the question of how sensitive posterior inference for $\boldsymbol \alpha$ can be to prior specification and the compatibility between the expert opinions and the observed data.
In this section we study these questions in simple settings using simulations.
We employ four hyperpriors in our analyses: a $\text{Dirichlet}(1, 1, 1, 1, 1)$ and a more flexible $\text{Dirichlet}(1/10, 1/10, 1/10, 1/10, 1/10)$, along with the corresponding moment-matching logistic-normal priors.
All computations are done with the simple Monte Carlo method outlined in the end of section~\ref{sec:hierPrior} and agree closely with solutions using MCMC (not shown).

Central to the discussion here is the notion that the marginal likelihoods -- of each expert prior--, when suitably normalised, provide a gold standard for the weights, in the sense that one could do no better when learning the weights conditional on the observed data.
We shall refer to the vector of weights obtained by dividing the marginal (integrated) likelihood of each expert by the sum of marginal likelihoods by $\boldsymbol\alpha^{\prime\prime}$.
If each prior were a model for the data, then $\boldsymbol\alpha^{\prime\prime}$ would be the model probabilities or the Bayesian model averaging (BMA) weights.

The examples explored here are highly stylised and the marginal likelihood is not usually available for comparison.
As such, the discussion presented here should be seen as the analysis of a baseline scenario, where we have control over the ground truth and can more clearly analyse the issues of estimating and interpreting the weights in logarithmic pooling.
Note also that the goal of this section is not to address the long run (frequentist) properties of the posterior distribution of the weights -- we provide the results of a brief experiment addressing some aspects of posterior concentration under repeated sampling in Figure~\ref{fig:concentration_results_normal}.
Rather, the goal of the following sections is to provide some insight into the potential pitfalls of the interpretation of the posterior weights even in the context of simple models.

\subsubsection{Example 1: Beta conjugate analysis}
\label{sec:learning_rate_beta}

For our first example in this section, we will use $K = 5$ experts, who will elicit Beta distributions about a probability $p$.
Some data $(x, n)$ will then be observed and a likelihood $L(x \mid n, p) = \text{binomial}(n, p)$ will summarise the information brought by the data.
In what follows we will elicit the parameters of a Beta distribution on $p$ for each expert using the mean $\mu_i :=  \mathbb{E}_i[p]$ and coefficient of variation $c_i := \sqrt{\text{Var}_i(p)}/ \mu_i$.
For more information, please see the appendix of~\cite{Coelho2015}.

The
setting we will investigate is when one expert provides a distribution that is significantly more compatible with the data that are ultimately observed.
The idea here is to evaluate how the posterior distribution of the weights $p(\boldsymbol\alpha \mid x, n)$ supports the ``correct'' expect as we vary (a) the strength of evidence $x/n$ and (b) the coefficient of variation of the ``correct'' distribution.
As we make the correct expert's coefficient of variation smaller, we expect the posterior weight to increase.
The intuition is that if one gives the correct answer with more certainty, one should receive more credence~\textit{a posteriori}.
To study point (a), we evaluate the posterior weights for $x/n = \{5/10, 50/100, 500/1000, 5000/10000 \}$.
In addition, we choose the ``true'' $p = 1/2$ and then construct $\boldsymbol{m} = \{0.1, 0.2, 0.5, 0.8, 0.9\}$ and $\boldsymbol c = \{0.1, 0.1, c_2, 0.1, 0.1 \}$, where we will vary $c_2$ between $0.001$ and $0.75$ in order to study point (b) above.
The upper value was chosen such that this is the largest value of $c_2$ for which the weight of the ``correct'' expert in $\boldsymbol\alpha^{\prime\prime}$ is the highest weight when $x/n=5/10$.

\begin{figure}[!ht]
\begin{center}
\subfigure[][Ratios of marginal likelihoods]{\includegraphics[scale=.45]{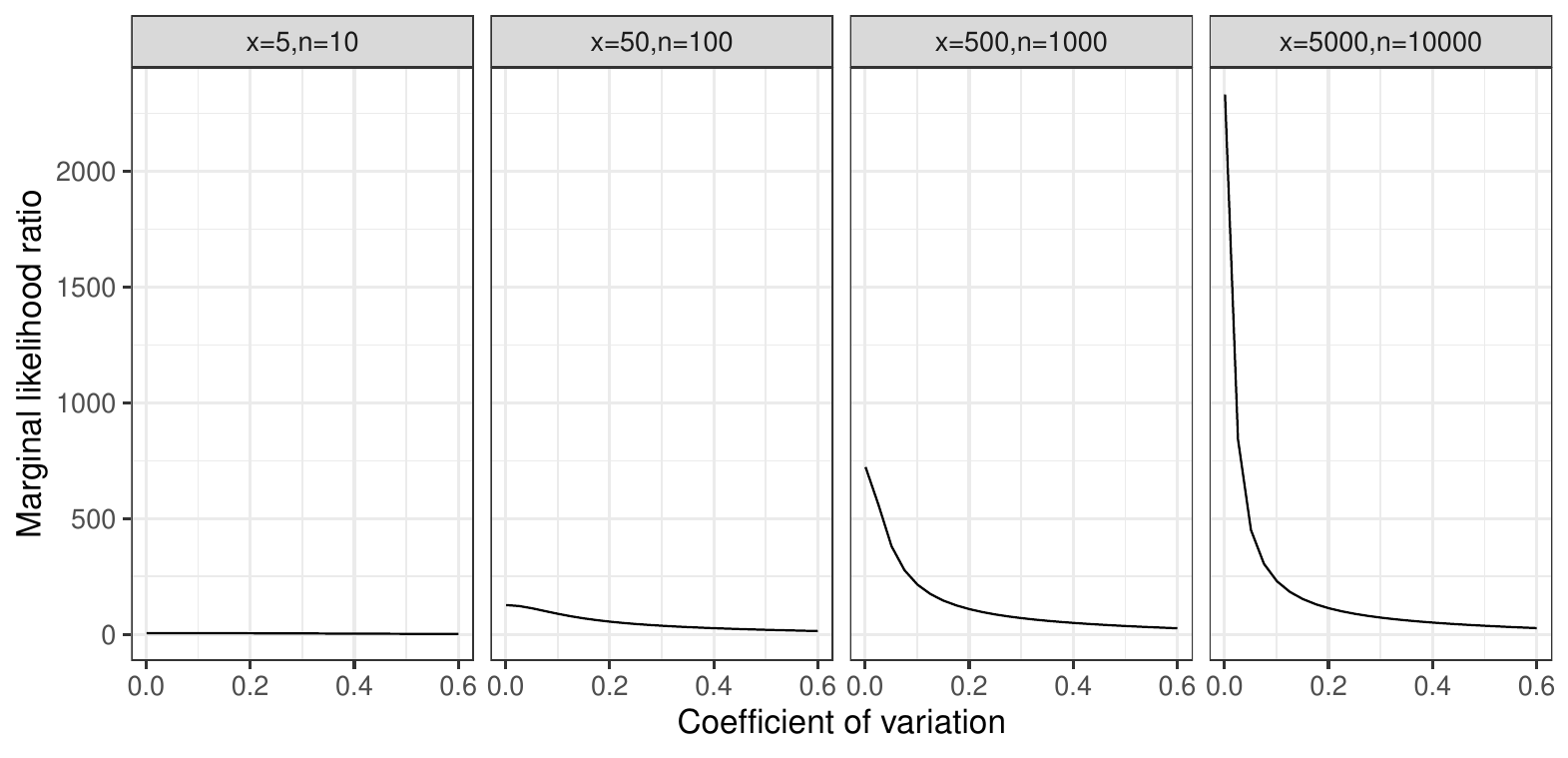}}
\subfigure[][Ratios of posterior weights]{\includegraphics[scale=.45]{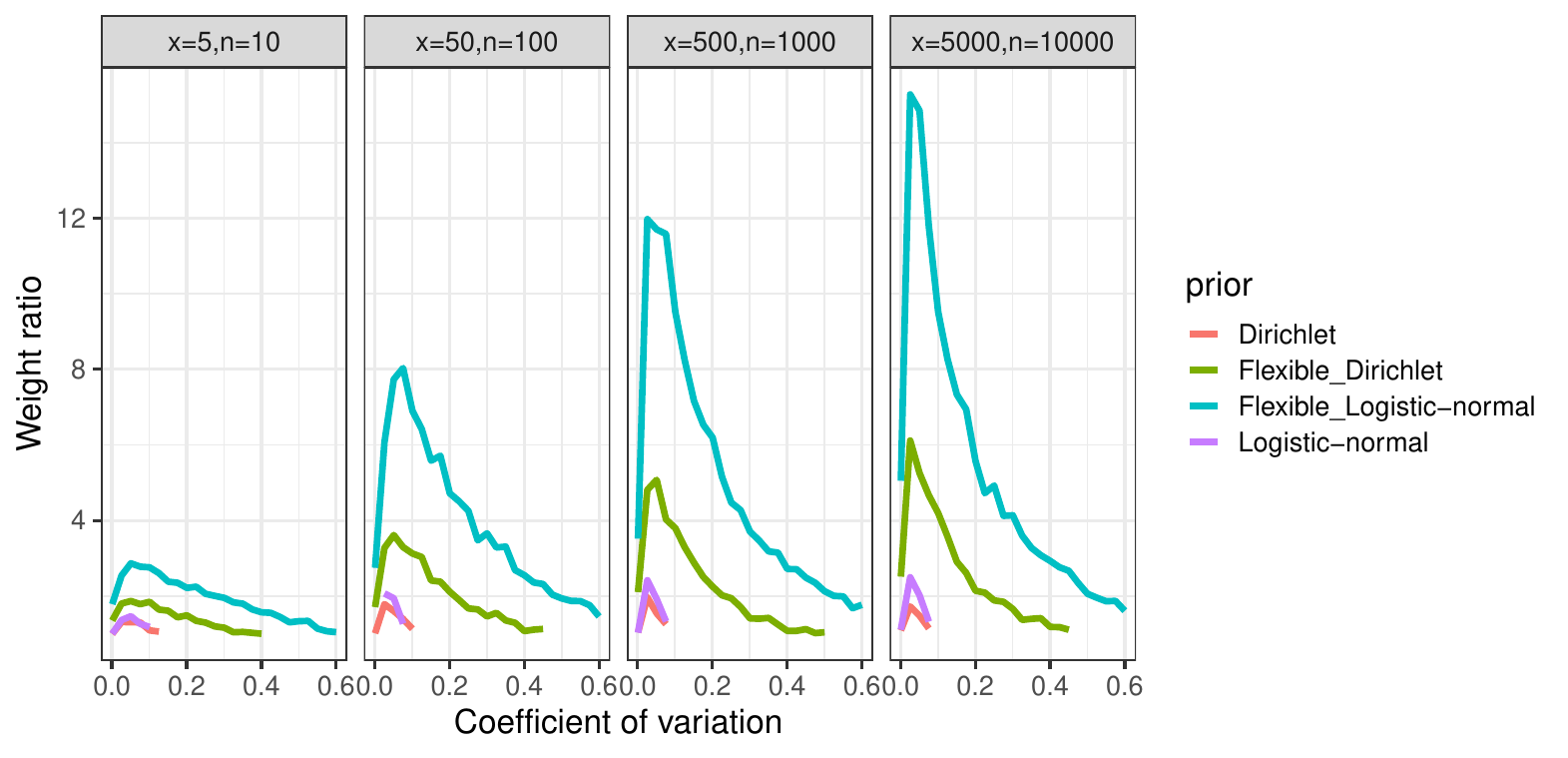}}
\end{center}
\caption{\textbf{Marginal likelihood and weight ratios for the simulated situation with one correct expert, various strengths of evidence}.
Panel (a) shows the ratio between the largest and second largest marginal likelihoods ($r_l$) as the correct expert's coefficient of variation ($c_2$) changes, while panel (b) shows the ratio between the largest and second largest posterior mean weights ($r_w$) in the same settings.
Vertical tiles show the observed data and colours in panel (b) show the hyperprior on $\boldsymbol\alpha$.
``Flexible'' priors are a Dirichlet(1/10, 1/10, 1/10, 1/10, 1/10) and the corresponding moment-matching logistic-normal.
We interrupt the lines for values of $c_2$ for which expert 2, the correct expert, does not attain the largest posterior weight (see text). 
}
\label{fig:one_correct_results_beta}
\end{figure}

In Figure~\ref{fig:one_correct_results_beta}, we plot two quantities as a function of the coefficient of variation of the correct expert ($c_2$) for various levels of evidence $(x, n)$: (i) the ratio between the largest and second largest marginal likelihoods ($r_l$)  computed using (\ref{eq:marginalbeta}); and (ii) the ratio between the posterior means of the largest and second largest weights ($r_w$).
The marginal likelihoods (Figure~\ref{fig:one_correct_results_beta}a) behave as expected, with $r_l$ diminishing as $c_2$ increases, the effect more pronounced with increasing the strength of evidence ($x/n$).
We show $r_w$ as function of $c_2$ in Figure~\ref{fig:one_correct_results_beta}b.
While for larger values of $c_2$ the ratio decreases as expected, for low values (high precision) it is also low, attaining a maximum at an intermediate value.
This somewhat counter-intuitive result is a quirk of Beta distributions: for low values of $c_2$, the corresponding parameter values $a_2 = b_2$ are large compared to other $(a_i, b_i)$, which means that any configuration of the weights that assigns non-zero weight to expert 2 is likely to lead to a combined prior that is compatible with the data $x/n$.

If $c_2$ is large, the ``correct'' distribution becomes too diffuse and a different expert is favoured.
To convey this in Figure~\ref{fig:one_correct_results_beta}b, we interrupt the plotted lines for values of $c_2$ at which expert 2 was not the one with the highest weight.
The correct expert does not attain the largest posterior weight for all values of $c_2$ for three of the four hierarchical priors considered. 
The ``flexible'' logistic-normal prior is the only hyperprior for which expert 2 is consistently favoured for all values of $c_2$ considered.
This phenomenon is similar in nature to identifiability issues in linear mixtures, where components need to be well separated in order for it be possible to reliably recover the mixing proportions~\citep{Yakowitz1968}. 
The results show that the ``flexible'' logistic-normal hyperprior, i.e., a moment-matching prior to the $\text{Dirichlet}(1/10, 1/10, 1/10, 1/10, 1/10)$, circumvents these identifiability problems and allows for better discrimination of the ``correct'' expert.

The results so far make clear that interpreting the posterior distribution of the weights, in particular the posterior means, is not necessarily trivial or intuitive.
We give an explicit example to illustrate the inherent problem of interpreting the weights in a log-linear mixture of beta distributions.
Suppose $c_2 = 0.2$ and $c_j = 0.1$ for all $j \neq 2$, with $\boldsymbol{m}$ given as before.
This setup leads to $\boldsymbol a = \{ 89.9, 79.8, 12.0, 19.2, 9.1\}$ and $\boldsymbol b = \{809.1, 319.2, 12.0, 4.8, 1.01\} $.
If the data are $x = 5$ and $n = 10$, computing marginal likelihoods and normalising would lead to weights $\boldsymbol\alpha^{\prime\prime} = \{0.006, 0.095, 0.710, 0.142, 0.048\}$.
However, by calculating $a^{\star\star} = \sum_{i = 0}^K \alpha_i^{\prime\prime} a_i = 19.75$ and $b^{\star\star} =  \sum_{i = 0}^K \alpha_i^{\prime\prime} b_i = 44.00$, we see that we obtain a pooled prior with $\mathbb{E}_\pi [p] =  0.31$, far off the ``optimal'' $1/2$.
Even in this situation where $r_l \approx 5$, weighting experts according to their marginal likelihoods does not lead to a satisfactory solution.
Hence, we argue that there is no hope to reliably learn the weights from these data under this configuration of the expert opinions.
If the data were, say, $x = 50, n = 100$, then one would obtain marginal likelihood-based weights such that the pooled ``prior'' expectation would be $\mathbb{E}_\pi [p] = 0.51$.

\subsubsection{Example 2: Gaussian conjugate analysis with known variance}
\label{sec:learning_rate_Gaussian}

Next, we explore an example where the mean and coefficient of variation setup above is easier to interpret.
Consider the problem of drawing inference about the mean $\mu$ of a Gaussian distribution with known variance $\sigma^2$.
Similar to the example above, we investigate the estimation of posterior weights under various scenarios for the expert opinions.
For this set of experiments we generated $n=10, 100, 500$ data points from a Gaussian distribution with $\mu=3$ and  $\sigma^2 = 1$.
We then considered $\boldsymbol{m} = \{1, 2, 3, 4, 5\}$ and $\boldsymbol c = \{0.1, 0.1, c_2, 0.1, 0.1 \}$, where we vary $c_2$ between $0.001$ and $1.5$, with this upper bound being chosen similarly to what was done above, in this case for the data set with $n=10$.

Results (see Figure~\ref{fig:one_correct_results_normal}) show the same pattern as Figure~\ref{fig:one_correct_results_beta}, with the ratio of weights being small for very low cvs and then increasing with cv until it starts to decay, as expected.
This phenomenon is rooted in the same numerical cause as what is observed for the Beta example; for very low cvs, the variance for expert 2 is really small, which in turn makes $w_2 = \alpha_2/v_2$ large for pretty much any value of $\alpha_2$.
This in turn means that many weight configurations $\boldsymbol{\alpha}$ will lead to very similar values of the pooled hyperparameters, $m^\ast$ and $v^\ast$ and thus will not receive very different probability~\textit{a posteriori}.

\subsection{Bayesian melding with varying weights}
\label{sec:melding_apps}

We now turn our attention to applications of logarithmic pooling to the statistical analysis of deterministic models.
In their seminal paper,~\cite{Poole2000} lay out Bayesian melding as way to achieve full Bayesian inference for deterministic models -- see also Section~\ref{sec:background_melding} above.
In this section we explore two Bayesian melding applications and extend their approach by accommodating uncertainty about the weight $\alpha$. 

\subsubsection{Bowhead whale population growth}
\label{sec:bowhead}

We begin with the analysis of a non-age-structured population deterministic model (PDM) population model for bowhead whales originally carried out by~\cite{Poole2000}.
The model describes the annual population of bowhead whales in terms of the annual number of whales killed , $C_t$, the maximum sustainable yield rate (MSYR) and the initial bowhead population ($P_0$) as:
\begin{equation}
\label{eq:popmodel_bowhead}
 P_{t + 1} = P_t - C_t \times \text{MSYR} \times P_t \left( 1- (P_t/P_0)^2 \right).
\end{equation}
One of the quantities of interest in the model was $P_{\text{1993}}$, due to 1993 being the last year for which independent abundance measurements were available, allowing for model calibration.
Another important model quantity is the rate of population increase from 1978 to 1993, ROI, defined through
\begin{equation*}
 \label{eq:ROI_P1993}
P_{1993} = P_{1978}(1 + \text{ROI})^{15}. 
\end{equation*}
We are then interested in the model outputs $\phi = \{P_{1993}, \text{ROI}\}$.
The key idea is to account for the influence of the priors on the inputs $\theta = \{\text{MSYR}, P_0\}$ on $P_{1993}$ through the~\textbf{induced} distribution.
In particular, we aim at composing the prior distribution
\begin{equation}
 \label{eq:P1993_pool}
 \tilde{q}_{\Phi}(P_{1993}) \propto q_1^\ast(P_{1993})^\alpha q_2(P_{1993})^{1-\alpha},
\end{equation}
where $q_1^\ast$ is the induced distribution and $q_2$ is the natural prior on $P_{1993}$.
The main innovation we propose here is to place a probability distribution over $\alpha$ in order to relax the need to fix it to particular value.
We choose a $\text{Beta}(1, 1)$ prior as our $\pi_A$.
The target posterior is then 
\begin{equation}
 \label{eq:bowhead_posterior}
  p_{\Theta, M}(P_0, \text{MSYR}, \alpha \mid C_t) \propto \tilde{q}_{\Theta}(P_0, \text{MSYR}) L_1(P_0, \text{MSYR}) L_2(P_{1993})\pi_A(\alpha),
\end{equation}
where $\tilde{q}_{\Theta}$ is the suitably inverted distribution over the input space from the prior over the output space, $\tilde{q}_{\Phi}$ (see~\cite{Poole2000}, section 3.3.4). 
The subscript makes reference to the fact that this is a posterior over the inputs $\theta \in \Theta$ which are linked to the outputs $\phi \in \Phi$ by a deterministic model $M$, given by~(\ref{eq:popmodel_bowhead}).
Further details on priors and likelihoods are given in~\cite{Poole2000} and the Appendix of this paper.
We note that when $\alpha$ is random, it is important to include all of the normalising constants that depend on it~\citep{Neuenschwander2009}, in particular the normalising constant of the expression in (\ref{eq:P1993_pool}).

Here we will consider two ways of approximating (\ref{eq:bowhead_posterior}).
First, we used the sampling importance-resampling (SpIR) algorithm described in Appendix~\ref{sec:spIR}.
This method does not rely on any parametric approximation to the induced distribution $q_1^\ast$, instead using standard kernel methods to approximate the density at any point.
We used $k = l = 100, 000$ iterations to produce a sample from $p_{\Theta, M}$.
We also explored a Hamiltonian Monte Carlo (HMC) implementation in Stan~\citep{Carpenter2017}.
However, for this implementation we needed to approximate $q_1^\ast$ by a parametric form.
Since $q_2$ is a normal distribution, we approximate $q_1^\ast$ by a normal distribution such that $\tilde{q}_{\Phi}$ (Equation~\ref{eq:P1993_pool}) can be written in closed-form.
We give further discussion on this choice in Appendix~\ref{sec:appendix_bowhead}.
Since $p_{\Theta, M}$ is a challenging target distribution, we used four independent chains of  $10,000$ iterations each.
We observed a low percentage of divergent iterations ($<$2\%), likely caused by the very challenging posterior geometry induced by high correlations between parameters.

\begin{figure}[!ht]
\begin{center}
\includegraphics[scale=.45]{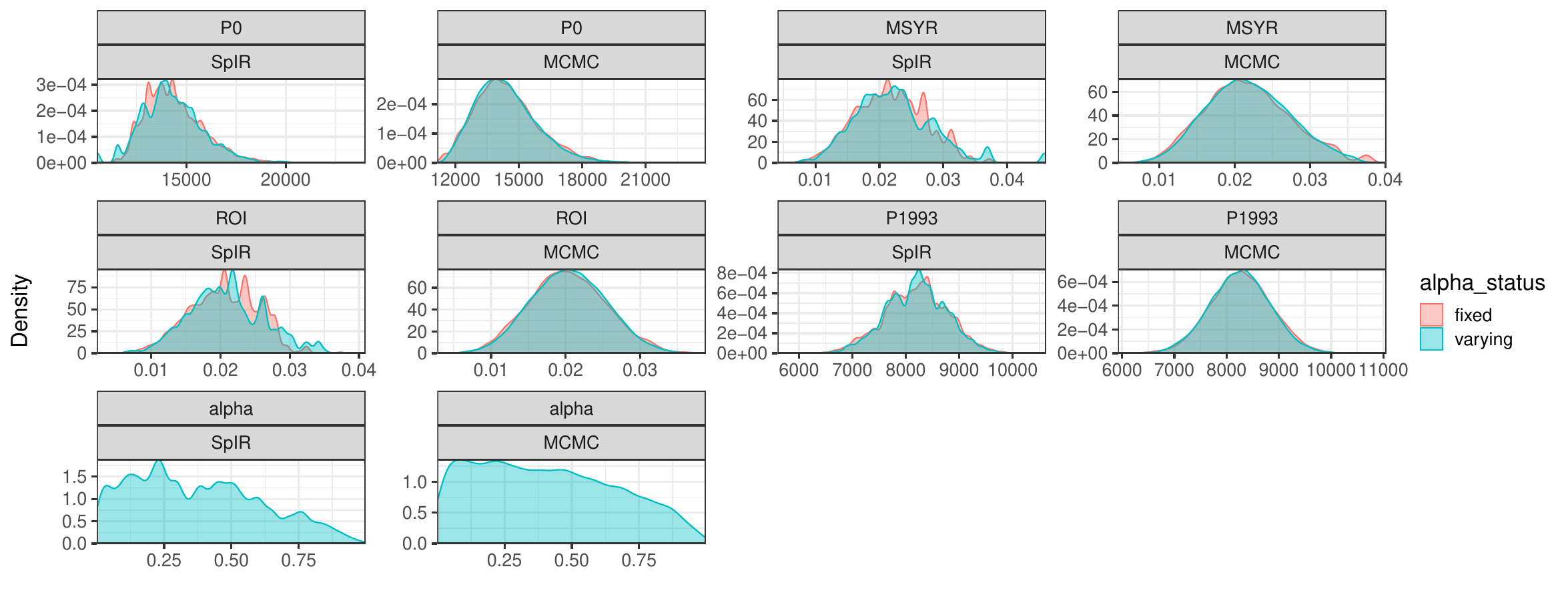}
\end{center}
\caption{\textbf{Marginal posterior distributions for various quantities of interest in the bowhead population model}.
We show the posterior distributions obtained by using sampling importance-resampling (SpIR) and Markov chain Monte Carlo (HMC-MCMC), for fixed $\alpha = 1/2$ and placing a prior $\pi_A$ on $\alpha$ (``varying'').
}
\label{fig:bowhead_marginal_posteriors}
\end{figure}

In Figure~\ref{fig:bowhead_marginal_posteriors} we show the marginal posteriors for various quantities of interest, obtained with both algorithms and for fixed and varying $\alpha$.
As expected, SpIR are a bit noisier, but distributions are largely the same as obtained by MCMC.
For $\alpha$ in particular, despite the ruggedness of distribution obtained with SpIR, the mean and 95\% credibility intervals of both distributions match very closely: SpIR = 0.39 (0.02--0.87) and MCMC = 0.40 (0.02--0.91).
The high posterior uncertainty about $\alpha$ and the substantial overlap between distributions with fixed and varying $\alpha$ could be explained by the lack of sensitivity of the posterior distribution to $\alpha$.
We confirm this is indeed the case by running SpIR (original algorithm by~\cite{Poole2000}) for a few values of $\alpha$ (including the endpoints $0$ and $1$) and verifying very little difference in the resulting posteriors (Figure~\ref{sfig:alpha_sensitivity_bowhead}). 

\subsubsection{Influenza in a boarding school}
\label{sec:SIR_flu}

Another important class of deterministic models are the ordinary differential equation-based models of disease transmission.
Here we will consider such a deterministic epidemic model and how one can draw inference about a key epidemiological quantity, the basic reproductive number, $R_0$.
In 1978, an anonymous source reported an influenza H1N1 epidemic at a small boarding school in England~\citep{Anonymous1978}.
In total, 512 boys out of 763 became ill during the outbreak.
Due to the population being isolated and having high rates of contact, many of the assumptions of compartimental epidemic models hold.
In particular, the Susceptible-Infected-Removed (SIR) model is a good description of disease spread.
The model consists of the system of ordinary differential equations
\begin{eqnarray*}
\frac{dS}{dt}&=& - \beta SI,\\
\frac{dI}{dt}&=&  \beta SI - \gamma I,\\
\frac{dR}{dt}&=& \gamma I, 
\end{eqnarray*} 
where  $S(t) + I(t) + R(t) = 1 \: \forall t$, $\beta$ is the transmission (infection) rate and $\gamma$ is the recovery rate.
The basic reproductive number is 
\begin{equation}
\label{eq:r0def}
R_0 = \frac{\beta}{\gamma}. 
\end{equation}
The goal is to draw inference about $\beta$ and $\gamma$, and consequently about $R_0$, from data.
Data on the number of infected individuals per time ($Y(t)$) were obtained from the~\textbf{outbreaks} package~\citep{Outbreaks2019} and we choose to model the deviation from the ODE solution using log-normal errors, i.e.,
\begin{equation}
 \label{eq:log-normal_likelihood}
 L(Y(t)\mid \beta, \gamma, \sigma_I^2) = \text{log-normal}(\mu =  \log(I(t)), \sigma_I^2),
\end{equation}
where $I(t)$ is computed~\textit{via} an ODE solver.
Here we will consider a situation where one has priors on $\beta$ and $\gamma$, which induce a prior $q_1^\ast$ on $R_0$, and also a prior $q_2$ on $R_0$ directly.
This is the case when, for instance, one wants to make $q_2$ informative so as to incorporate expert knowledge and/or evidence from previous study.
For the priors on $\beta$ and $\alpha$ we choose commonly used, so-called ``uninformative'' log-normal priors with parameters $\mu_{\beta} = \mu_{\gamma} = 0$ and $\sigma_{\beta}^2 = \sigma_{\gamma}^2 = 1$, which induces a log-normal distribution ($q_1^\ast$) on $R_0$ with parameters $\mu_1 = \mu_\beta - \mu_\gamma$ and $\sigma_1^2 = \sigma_{\beta}^2 +  \sigma_{\gamma}^2$.
Using the extensive information gathered by~\cite{Biggerstaff2014}, we constructed an informative log-normal prior ($q_2$) with mean $1.5$ and variance $0.25^2$, which gives $\mu_2 = 0.3917656$ and variance $\sigma_2^2 =  0.1655264$.
This leads to a prior credibility interval of ($1.070$--$2.047$), which covers most of the estimates (and confidence intervals) of $R_0$ for Influenza found by~\cite{Biggerstaff2014}.
The target posterior is then
\begin{equation}
 \label{eq:target_SIR}
 p(\beta, \gamma, \alpha \mid Y(t)) \propto  L(Y(t)\mid \beta, \gamma, \sigma_I^2) q_1^\ast(R_0)^\alpha q_2(R_0)^{1-\alpha}\pi_A(\alpha),
\end{equation}
where we again let $\pi_A$ be a Beta(1,1) distribution.
This setup is convenient because it leads to a closed-form expression for the combined prior on $R_0$ (see Appendix, Section~\ref{sec:appendix_common_poolings}), while the log-normal priors are flexible and useful in practice.
We approximate the posterior in (\ref{eq:target_SIR}) using HMC as described in Appendix~\ref{sec:appendix_compdetails}.

In Figure~\ref{fig:SIR_results}a we show the posterior distribution of the pooling weight $\alpha$, which favours high values with a mean and 95\% credibility interval of 0.77 (0.21--0.99).
The posterior distribution for $R_0$ obtained by letting $\alpha$ vary and also the resulting distributions of fixing $\alpha = 1/2$ or $\alpha = 1$ are shown in Figure~\ref{fig:SIR_results}b.
One can see that fixing $\alpha = 1$ and hence excluding the informative prior leads to a higher estimate of $R_0$ and fixing $\alpha = 1/2$ as per~\cite{Poole2000} leads to the lowest estimates.
The solution proposed in this paper, namely assigning $\alpha$ a prior and estimating it from data, leads to an intermediate solution.
Fixing $\alpha = 1/2$ also leads to underestimating the measured incidence (Figure~\ref{fig:SIR_results}c), whilst setting $\alpha = 1$ leads to mean predictions that are higher, albeit still underestimating the measured incidence.
Again, letting $\alpha$ vary leads to an intermediate solution.

\begin{figure}[!ht]
\begin{center}
\subfigure[][$\alpha$]{\includegraphics[scale=.35]{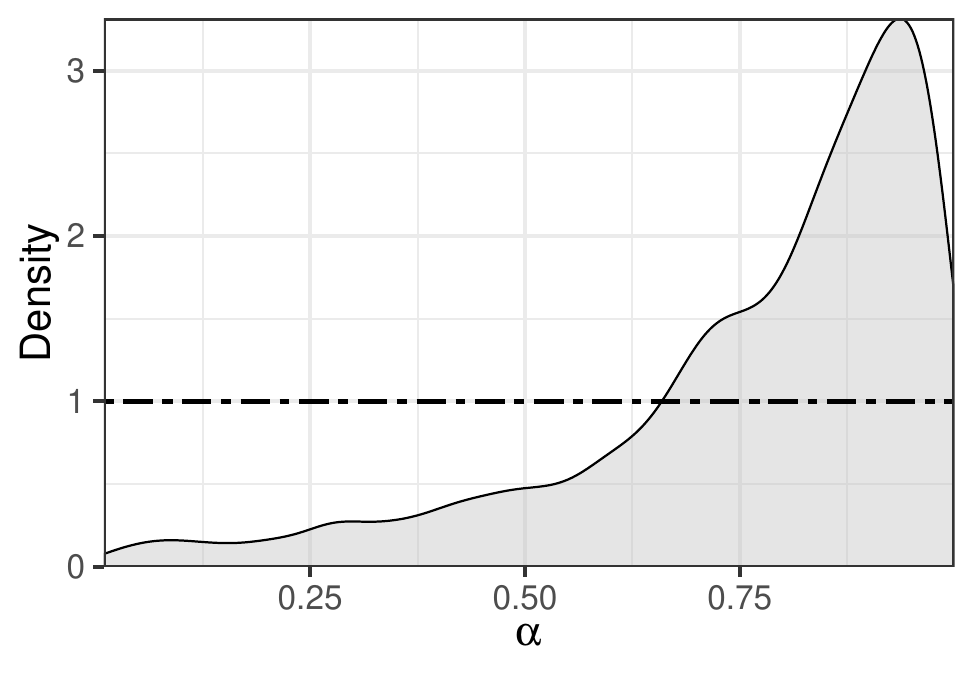}}
\subfigure[][$R_0$]{\includegraphics[scale=.35]{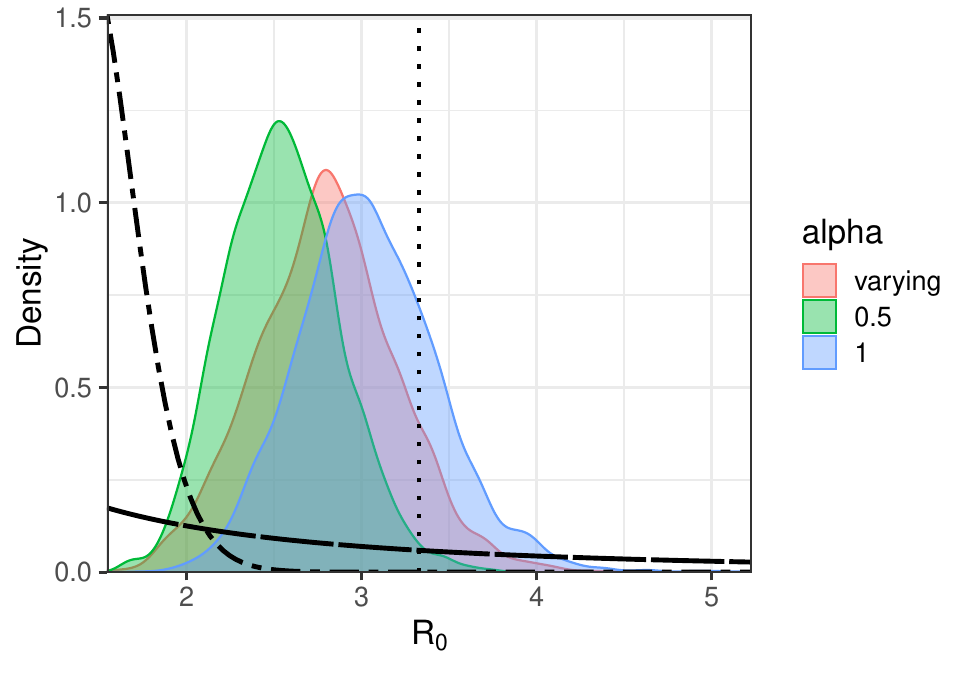}}
\subfigure[][Predictions]{\includegraphics[scale=.45]{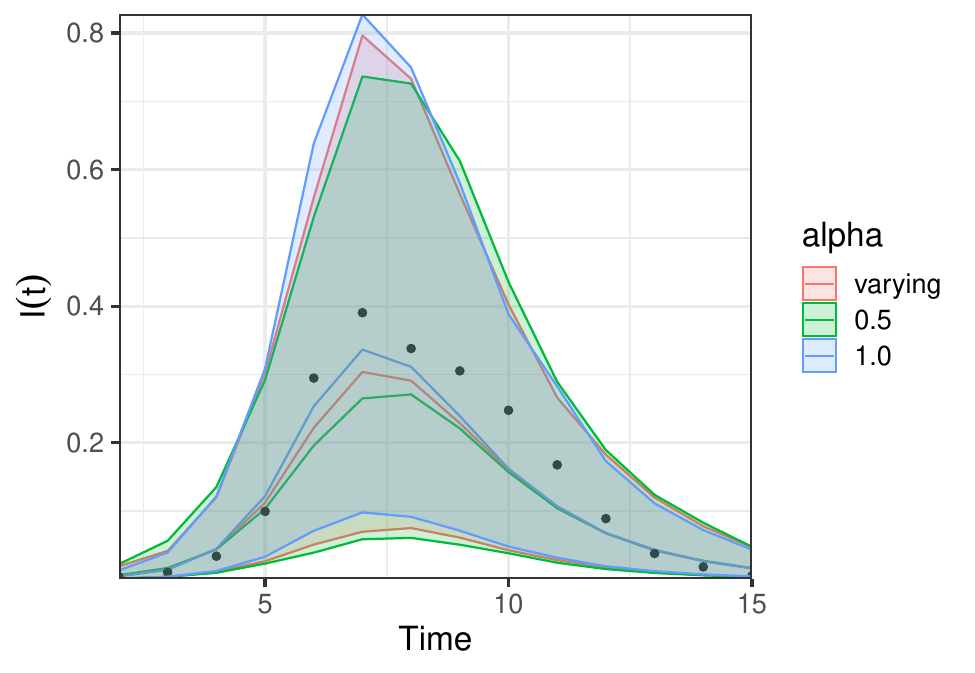}}

\end{center}
\caption{\textbf{Estimates of the pooling weight ($\alpha$), the basic reproductive number ($R_0$) and predictions of the number of infected individuals}.
The posterior distribution for the pooling weight $\alpha$ is shown in panel (a), where the horizontal dashed line shows the prior density, a Beta(1, 1).
Panel (b) shows the posterior distribution for $R_0$ obtained with estimating $\alpha$ (``varying'') or fixing it to either $1/2$ or $1$.
Vertical line shows $R_0 = 3.78$~\citep{Murray2002}, the dot-dashed line shows the informative prior $q_2$ and the longdash line shows the induced prior $q_1^\ast$.
In panel (c) we show the posterior mean and 95\% credibility intervals for the proportion of infected individuals, again by either letting $\alpha$ vary or fixing it to either $1/2$ or $1$.
}
\label{fig:SIR_results}
\end{figure}

Our results agree somewhat with the estimate obtained by~\cite{Murray2002}, who finds $\rho = N/R_0 = 202$ and hence $R_0 = 3.78$, using purely numerical methods with no acknowledgment of uncertainty.
The highest estimates we obtained were for fixed $\alpha = 1$, $R_0$ = 3.02 (2.27--3.83).
This example showcases a desirable consequence of letting $\alpha$ vary: when the ``natural'' prior $q_2$ -- which is normally informative -- is incompatible with the data, it will receive a lower weight ($\alpha$ closer 1) and hence allow the induced prior ($q_1$), which is usually more diffuse, to dominate.
In fact, as discussed by~\cite{Biggerstaff2014}, the spread of the 1978 boarding school epidemic is unusually fast when compared to regular seasonal Influenza and was likely caused by the lack of previous exposure of the population to the causing strain, H1N1.
The varying $\alpha$ approach makes it possible to deal with such an outlier data set by lowering the influence of the informative prior constructed based on previous studies.

\section{Discussion}
\label{sec:discussion}

In this paper we have provided an overview of statistical applications of logarithmic pooling (LP), including a new approach based on assigning a prior measure to the weights.
In what follows we discuss our findings in light of the rich literature on log-pooling, as well as point out connections to other parts of the statistical literature on model and forecast aggregation.

\subsection{Objections to logarithmic pooling and their counter-arguments}

\cite{West1984} argues that LP is strictly theoretically justifiable only when the expert opinions agree.
Moreover, LP also violates basic coherence in other respects, for instance when one considers marginalisation or other probability manipulations. 
\cite{Genest1986B} (pg. 124) explain, however, that these conclusions stem from the restrictive assumption that the group utility is expressed as a function of the individual utilities.
In a statistical application context, the expert opinions are usually employed by an independent decision maker, henceforth called the analyst, and she has her own utility function which can be assumed to not depend on the individual utilities.
Another quirk of logarithmic pools is that $\supp(\pi) = \cap_{i=0}^K \supp(f_i)$, i.e., the pooled distribution will have the smallest support amongst the distributions being combined.
This means a single expert can make large portions of the sample space impossible under the pooled distribution.
Again, however, the analyst can use external considerations to exclude an expert whose probability density has too narrow a support.

A consequence of encoding opinions as probability densities is that representations of the same information might have different properties depending on the choice of dominating measure.
The results of meta analysis in Section~\ref{sec:metaAnalysis} make this clear: choosing to represent the information brought by the studies as Beta distribution or a Gaussian does not affect the numerical values of means and probability intervals, but does seem to impact the optimality-based methods for choosing the weights, in particular minimum KL  (Table~\ref{tab:weights_MSM}).
On the other hand, as shown by the agreement of the probability intervals in the bottom of Table~\ref{tab:prior_MSM}, moving away from optimality criteria and instead assigning a prior distribution to the weights largely removes dependence on specific choices of probability densities by properly accommodating uncertainty about the weights.

One might worry about being able to learn the weights from data, since the weights depend on the likelihood only indirectly.
Indeed, as shown in Section~\ref{sec:learning_rate}, it might not always be possible to identify the expert whose opinion is most consistent with the observed data.
We argue, however, that this happens mostly in situations where one would not expect to learn much anyway.
Consider the situation where the data are highly informative (e.g. Gaussian with small known variance): the likelihood will dominate the prior in the posterior, meaning that most weight configurations will lead to similar (joint) posterior densities.
This happens unless there is substantial disagreement between the individual priors being combined and we argue it is a desirable property of LP.

As a final caveat, we note that if interest lies on a multivariate quantity $\theta \in \mathbb{R}^d$, $d>1$, obtaining the normalising constant $t(\boldsymbol\alpha)$ will entail computing a high-dimensional integral, which is infeasible to do via quadrature.
Here, importance sampling techniques can be leveraged to provide stable and accurate estimates of normalising constants (see Future directions).

\subsection{The case for (hierarchical) logarithmic pooling}

We shall now argue that properties such as external Bayesianity, relative propensity consistency and log-concavity make logarithmic pooling a powerful tool for the analyst.

Mainly due to the simplicity of their construction, linear mixtures are much more popular in statistical applications~\citep{Fruhwirth2019} than their log-linear cousins.
As we hope to have shown in this paper, however, log-linear mixtures (logarithmic pooling) can be as useful or more.
External Bayesianity means one does not need to worry about combining the priors first and then obtaining the posterior; one can simply take a set of posterior distributions computed with the same likelihood and combine them.

Moreover, LP preserves log-concavity, which might be crucial in computationally demanding settings where slice sampling, variational or other algorithms that assume log-concavity are employed.
In summary, we argue that by employing logarithmic pooling to combine probability densities, the analyst is making the best use of the available information by forming a coherent distribution, that preserves many of the features encoded by the experts in their opinions.

After its theoretical properties, the strongest argument in favour of LP is by far is its adaptability.
The extra flexibility brought on by the hierarchical prior on the weights might prove crucial in scientific applications where decision under uncertainty is a regular occurrence.
For example, a main strength of Bayesian melding is downweighting parameter values based on implausible model outputs.
This strength is magnified by using a hierarchical prior that allows the weight parameter to vary.
Indeed,~\cite{Poole2000} (Section 5.2) argue that estimating $\alpha$ would be a fruitful path to explore and our results corroborate that view.
The result in Section~\ref{sec:SIR_flu} makes clear the potential of varying-weights Bayesian melding for resolving prior-data conflict.
In particular, for protecting the analyst from drawing strong conclusions when the ``natural'' prior on the quantity of interest is in disagreement with the information brought by the data under analysis.

When comparing the hierarchical prior approach to optimality-based procedures, one might argue that excluding a few or even all experts but one is not problematic since a few experts may, when suitably combined, summarise the information provided by the whole group.
Whilst the weights are not probabilities, we argue that it would be preferable to have a solution that respects the so-called Cromwell's rule~\citep[pg. 91]{Lindley2013}, i.e., not assigning zero probability to events that are logically possible.
Here this means allowing for the possibility that the opinion of all experts receives non-zero weight.
Incidentally, this should also help alleviate some of the problems discussed in the previous section.

\subsection{Future directions}

Future research will explore further applications of logarithmic pooling in statistical learning such as combining several posterior predictive distributions from different models fitted to the same data.
Techniques such as Bayesian predictive synthesis~(BPS, \cite{McAlinn2018, McAlinn2019a,McAlinn2019b}) and stacking~\citep{Yao2018} have focused on generalising linear pools to combine probabilistic predictions, and logarithmic pooling could be explored as possibility that preserves characteristics such as log-concavity and relative propensity consistency.
BPS can include logarithmic pooling as a special case, but understanding the conditions under which this holds remains an open question.

Another interesting avenue for the future is studying the interaction between variable transformations and logarithmic pooling.
An example is a situation where one has distributions about a probability $p$ but is interested in the log-odds, $\omega = \log(p/(1-p))$.
Should the experts be judged by how reasonable their distributions look in transformed space?
How to assign the weights in this situation?
 
In a practical setting, one might have a collection of MCMC (approximate) samples from different posterior distributions.
The statistical question then becomes how to sample from the pooled distribution by re-using these samples.
Such task would likely necessitate specially-designed MCMC methods, and would constitute a rich area of future inquiry.

In closing, we hope this paper (i) showcases the usefulness -- and potential pitfalls -- of logarithmic pooling as a way of combining probability distributions and (ii) entices the statistical community to add it to their toolbox.

\section*{Acknowledgments}

The authors would like to thank Professors Adrian Raftery, Christian Genest and Mike West, as well as  Drs. David Poole, Eduardo Mendes and Felipe Figueiredo for helpful suggestions.
DAMV and LSB were supported in part by CAPES under CAPES/Cofecub project (N. 833/15).
FCC is grateful to Funda\c{c}\~ao Getulio Vargas for funding during this 
project.
\bibliography{pooling}

\begin{thebibliography}{}

\bibitem[Abbas, 2009]{Abbas2009}
Abbas, A.~E. (2009).
\newblock A {K}ullback-{L}eibler view of linear and log-linear pools.
\newblock {\em Decision Analysis}, 6(1):25--37.

\bibitem[Aitchison and Shen, 1980]{Aitchson1980}
Aitchison, J. and Shen, S.~M. (1980).
\newblock Logistic-normal distributions: Some properties and uses.
\newblock {\em Biometrika}, 67(2):261--272.

\bibitem[Alkema et~al., 2008]{Alkema2008}
Alkema, L., Raftery, A.~E., and Brown, T. (2008).
\newblock Bayesian melding for estimating uncertainty in national hiv
  prevalence estimates.
\newblock {\em Sexually transmitted infections}, 84(Suppl 1):i11--i16.

\bibitem[Alkema et~al., 2007]{Alkema2007}
Alkema, L., Raftery, A.~E., Clark, S.~J., et~al. (2007).
\newblock Probabilistic projections of {HIV} prevalence using {B}ayesian
  melding.
\newblock {\em The Annals of Applied Statistics}, 1(1):229--248.

\bibitem[Anon., 1978]{Anonymous1978}
Anon. (1978).
\newblock Influenza in a boarding school.
\newblock {\em The British Medical Journal}, 1:587.

\bibitem[Bagnoli and Bergstrom, 2005]{Bagnoli2005}
Bagnoli, M. and Bergstrom, T. (2005).
\newblock Log-concave probability and its applications.
\newblock {\em Economic theory}, 26(2):445--469.

\bibitem[Barcellos et~al., 2003]{Barcellos2003}
Barcellos, N.~T., Fuchs, S.~C., and Fuchs, F.~D. (2003).
\newblock Prevalence of and risk factors for {HIV} infection in individuals
  testing for {HIV} at counseling centers in {Brazil}.
\newblock {\em Sexually transmitted diseases}, 30(2):166--173.

\bibitem[Betancourt, 2012]{Betancourt2012}
Betancourt, M. (2012).
\newblock Cruising the simplex: {H}amiltonian {M}onte {C}arlo and the
  {D}irichlet distribution.
\newblock In {\em AIP Conference Proceedings 31st}, volume 1443, pages
  157--164. AIP.

\bibitem[Betancourt, 2017]{Betancourt2017}
Betancourt, M. (2017).
\newblock A conceptual introduction to {H}amiltonian {M}onte {C}arlo.
\newblock {\em arXiv preprint arXiv:1701.02434}.

\bibitem[Biggerstaff et~al., 2014]{Biggerstaff2014}
Biggerstaff, M., Cauchemez, S., Reed, C., Gambhir, M., and Finelli, L. (2014).
\newblock Estimates of the reproduction number for seasonal, pandemic, and
  zoonotic influenza: a systematic review of the literature.
\newblock {\em BMC infectious diseases}, 14(1):480.

\bibitem[{Brazilian Ministry of Health}, 2000]{BMH2000}
{Brazilian Ministry of Health} (2000).
\newblock Bela vista \& horizon: Behavioral and serological studies among men
  who have sex with men.
\newblock Technical report, Brasília: Brazilian Ministry of Health.
\newblock In Portuguese.

\bibitem[Byrd et~al., 1995]{Byrd1995}
Byrd, R.~H., Lu, P., Nocedal, J., and Zhu, C. (1995).
\newblock A limited memory algorithm for bound constrained optimization.
\newblock {\em SIAM Journal on Scientific Computing}, 16(5):1190--1208.

\bibitem[Carneiro et~al., 2003]{Carneiro2003}
Carneiro, M., Cardoso, F.~A., Greco, M., Oliveira, E., Andrade, J., Greco,
  D.~B., and Antunes, C. M. d.~F. (2003).
\newblock Determinants of human immunodeficiency virus ({HIV}) prevalence in
  homosexual and bisexual men screened for admission to a cohort study of {HIV}
  negatives in {Belo Horizonte}, {Brazil: Project Horizonte}.
\newblock {\em Mem{\'o}rias do Instituto Oswaldo Cruz}, 98(3):325--329.

\bibitem[Carpenter et~al., 2017]{Carpenter2017}
Carpenter, B., Gelman, A., Hoffman, M.~D., Lee, D., Goodrich, B., Betancourt,
  M., Brubaker, M., Guo, J., Li, P., and Riddell, A. (2017).
\newblock Stan: A probabilistic programming language.
\newblock {\em Journal of statistical software}, 76(1).

\bibitem[Coelho and Carvalho, 2015]{Coelho2015}
Coelho, F.~C. and Carvalho, L.~M. (2015).
\newblock Estimating the attack ratio of dengue epidemics under time-varying
  force of infection using aggregated notification data.
\newblock {\em Scientific reports}, 5:18455.

\bibitem[Coelho and Code\c{c}o, 2009]{Coelho2009}
Coelho, F.~C. and Code\c{c}o, C.~T. (2009).
\newblock {{D}ynamic modeling of vaccinating behavior as a function of
  individual beliefs}.
\newblock {\em PLoS Comput. Biol.}, 5(7):e1000425.

\bibitem[Diaconis and Ylvisaker, 1979]{Diaconis1979}
Diaconis, P. and Ylvisaker, D. (1979).
\newblock Conjugate priors for exponential families.
\newblock {\em The Annals of Statistics}, pages 269--281.

\bibitem[Evans and Boersma, 1988]{Evans1988}
Evans, R.~J. and Boersma, J. (1988).
\newblock The entropy of a poisson distribution (c. robert appledorn).
\newblock {\em SIAM Review}, 30(2):314--317.

\bibitem[French, 1985]{French1985}
French, S. (1985).
\newblock Group consensus probability distributions: A critical survey in
  bayesian statistics.
\newblock {\em Bayesian statistics}, 2.

\bibitem[Fruhwirth-Schnatter et~al., 2019]{Fruhwirth2019}
Fruhwirth-Schnatter, S., Celeux, G., and Robert, C.~P. (2019).
\newblock {\em Handbook of mixture analysis}.
\newblock CRC press.

\bibitem[Gelman and Rubin, 1992]{Gelman1992}
Gelman, A. and Rubin, D.~B. (1992).
\newblock Inference from iterative simulation using multiple sequences.
\newblock {\em Statistical science}, pages 457--472.

\bibitem[Genest et~al., 1986]{Genest1986A}
Genest, C., McConway, K.~J., and Schervish, M.~J. (1986).
\newblock Characterization of externally bayesian pooling operators.
\newblock {\em The Annals of Statistics}, pages 487--501.

\bibitem[Genest et~al., 1984]{Genest1984}
Genest, C., Weerahandi, S., and Zidek, J.~V. (1984).
\newblock Aggregating opinions through logarithmic pooling.
\newblock {\em Theory and Decision}, 17(1):61--70.

\bibitem[Genest and Zidek, 1986]{Genest1986B}
Genest, C. and Zidek, J.~V. (1986).
\newblock Combining probability distributions: A critique and an annotated
  bibliography.
\newblock {\em Statistical Science}, pages 114--135.

\bibitem[Guardoni, 2002]{Guardoni2002}
Guardoni, G.~L. (2002).
\newblock On irrelevance of alternatives and opinion pooling.
\newblock {\em Brazilian Journal of Probability and Statistics}, pages 87--98.

\bibitem[Harrison et~al., 1999]{Harrison1999}
Harrison, L.~H., do~Lago, R.~F., Friedman, R.~K., Rodrigues, J., Santos, E.~M.,
  de~Melo, M.~F., Moulton, L.~H., Schechter, M., Group, P. P. O.~S., et~al.
  (1999).
\newblock Incident hiv infection in a high-risk, homosexual, male cohort in rio
  de janeiro, brazil.
\newblock {\em JAIDS Journal of Acquired Immune Deficiency Syndromes},
  21(5):408--412.

\bibitem[Hoffman and Gelman, 2014]{Hoffman2014}
Hoffman, M.~D. and Gelman, A. (2014).
\newblock The no-u-turn sampler: adaptively setting path lengths in
  {H}amiltonian {M}onte {C}arlo.
\newblock {\em Journal of Machine Learning Research}, 15(1):1593--1623.

\bibitem[{Jaynes}, 1957]{Jaynes1957}
{Jaynes}, E.~T. (1957).
\newblock {Information Theory and Statistical Mechanics. II}.
\newblock {\em Physical Review}, 108:171--190.

\bibitem[Jombart et~al., 2019]{Outbreaks2019}
Jombart, T., Frost, S., Nouvellet, P., Campbell, F., and Sudre, B. (2019).
\newblock {\em outbreaks: A Collection of Disease Outbreak Data}.
\newblock R package version 1.6.0.

\bibitem[Li et~al., 2017]{Li2017}
Li, Z.~S., Guo, J., Xiao, N.-C., and Huang, W. (2017).
\newblock Multiple priors integration for reliability estimation using the
  bayesian melding method.
\newblock In {\em Reliability and Maintainability Symposium (RAMS), 2017
  Annual}, pages 1--6. IEEE.

\bibitem[Lind and Nowak, 1988]{Lind1988}
Lind, N.~C. and Nowak, A.~S. (1988).
\newblock Pooling expert opinions on probability distributions.
\newblock {\em Journal of engineering mechanics}, 114(2):328--341.

\bibitem[Lindley, 2013]{Lindley2013}
Lindley, D.~V. (2013).
\newblock {\em Understanding uncertainty}.
\newblock John Wiley \& Sons.

\bibitem[Malta et~al., 2010]{Malta2010}
Malta, M., Magnanini, M.~M., Mello, M.~B., Pascom, A. R.~P., Linhares, Y., and
  Bastos, F.~I. (2010).
\newblock {HIV} prevalence among female sex workers, drug users and men who
  have sex with men in {Brazil}: a systematic review and meta-analysis.
\newblock {\em BMC Public Health}, 10(1):1.

\bibitem[McAlinn et~al., 2019]{McAlinn2019a}
McAlinn, K., Aastveit, K.~A., Nakajima, J., and West, M. (2019).
\newblock Multivariate bayesian predictive synthesis in macroeconomic
  forecasting.
\newblock {\em Journal of the American Statistical Association}.
\newblock arXiv:1711.01667. Published online: Oct 9 2019.

\bibitem[McAlinn et~al., 2018]{McAlinn2018}
McAlinn, K., Aastveit, K.~A., and West, M. (2018).
\newblock Bayesian predictive synthesis-- discussion of: Using stacking to
  average bayesian predictive distributions, by y. yao et al.
\newblock {\em Bayesian Analysis}, 13:971--973.

\bibitem[McAlinn and West, 2019]{McAlinn2019b}
McAlinn, K. and West, M. (2019).
\newblock Dynamic bayesian predictive synthesis in time series forecasting.
\newblock {\em Journal of Econometrics}, 210:155--169.
\newblock arXiv:1601.07463.

\bibitem[Murray, 2002]{Murray2002}
Murray, J.~D. (2002).
\newblock {\em Mathematical Biology I. An Introduction}, volume~17 of {\em
  Interdisciplinary Applied Mathematics}.
\newblock Springer, New York, 3 edition.

\bibitem[Myung et~al., 1996]{Myung1996}
Myung, I.~J., Ramamoorti, S., and Bailey~Jr, A.~D. (1996).
\newblock Maximum entropy aggregation of expert predictions.
\newblock {\em Management Science}, 42(10):1420--1436.

\bibitem[Neal et~al., 2011]{Neal2011}
Neal, R.~M. et~al. (2011).
\newblock Mcmc using {H}amiltonian dynamics.
\newblock {\em Handbook of markov chain monte carlo}, 2(11):2.

\bibitem[Neuenschwander et~al., 2009]{Neuenschwander2009}
Neuenschwander, B., Branson, M., and Spiegelhalter, D.~J. (2009).
\newblock A note on the power prior.
\newblock {\em Statistics in Medicine}, 28(28):3562--3566.

\bibitem[Pennock and Wellman, 1997]{Pennock1997}
Pennock, D.~M. and Wellman, M.~P. (1997).
\newblock Representing aggregate belief through the competitive equilibrium of
  a securities market.
\newblock In Geiger, D. and Shenoy, P.~P., editors, {\em Proceedings of the
  Thirteenth Conference on Uncertainty in Artificial Intelligence}, pages
  392--400. Morgan Kaufmann Publishers Inc.

\bibitem[Poole and Raftery, 2000]{Poole2000}
Poole, D. and Raftery, A.~E. (2000).
\newblock Inference for deterministic simulation models: the bayesian melding
  approach.
\newblock {\em Journal of the American Statistical Association},
  95(452):1244--1255.

\bibitem[{R Core Team}, 2019]{R2019}
{R Core Team} (2019).
\newblock {\em R: A Language and Environment for Statistical Computing}.
\newblock R Foundation for Statistical Computing, Vienna, Austria.

\bibitem[Raftery et~al., 2007]{Raftery2007}
Raftery, A.~E., Newton, M.~A., Satagopan, J.~M., and Krivitsky, P.~N. (2007).
\newblock Estimating the integrated likelihood via posterior simulation using
  the harmonic mean identity.
\newblock In Bernardo, J.~M., Bayarri, M.~J., Berger, J.~O., Dawid, A.~P.,
  Heckerman, D., Smith, A. F.~M., and West, M., editors, {\em {B}ayesian
  Statistics}, pages 1--45. Oxford University Press.

\bibitem[Rothman et~al., 2008]{Rothman2008}
Rothman, K.~J., Greenland, S., and Lash, T.~L. (2008).
\newblock {\em Modern Epidemiology}.
\newblock Lippincott Williams \& Wilkins, 3rd. edition.

\bibitem[Rufo et~al., 2012a]{Rufo2012A}
Rufo, M., Martin, J., P{\'e}rez, C., et~al. (2012a).
\newblock Log-linear pool to combine prior distributions: A suggestion for a
  calibration-based approach.
\newblock {\em Bayesian Analysis}, 7(2):411--438.

\bibitem[Rufo et~al., 2012b]{Rufo2012B}
Rufo, M.~J., P{\'e}rez, C.~J., Mart{\'\i}n, J., et~al. (2012b).
\newblock A bayesian approach to aggregate experts’ initial information.
\newblock {\em Electronic Journal of Statistics}, 6:2362--2382.

\bibitem[Savchuk and Martz, 1994]{Savchuk1994}
Savchuk, V.~P. and Martz, H.~F. (1994).
\newblock Bayes reliability estimation using multiple sources of prior
  information: binomial sampling.
\newblock {\em Reliability, IEEE Transactions on}, 43(1):138--144.

\bibitem[Sutmoller et~al., 2002]{Sutmoller2002}
Sutmoller, F., Penna, T.~L., de~Souza, C. T.~V., Lambert, J., Group, O. C. F.
  S.~P., et~al. (2002).
\newblock Human immunodeficiency virus incidence and risk behavior in the
  ‘{Projeto Rio}’: Results of the first 5 years of the {Rio de Janeiro}
  open cohort of homosexual and bisexual men, 1994--1998.
\newblock {\em International journal of infectious diseases}, 6(4):259--265.

\bibitem[Tun et~al., 2008]{Tun2008}
Tun, W., de~Mello, M., Pinho, A., Chinaglia, M., and Diaz, J. (2008).
\newblock Sexual risk behaviours and {HIV} seroprevalence among male sex
  workers who have sex with men and non-sex workers in {Campinas}, {Brazil}.
\newblock {\em Sexually Transmitted Infections}, 84(6):455--457.

\bibitem[West, 1984]{West1984}
West, M. (1984).
\newblock Bayesian aggregation.
\newblock {\em Journal of the Royal Statistical Society. Series A (General)},
  pages 600--607.

\bibitem[Yakowitz and Spragins, 1968]{Yakowitz1968}
Yakowitz, S.~J. and Spragins, J.~D. (1968).
\newblock On the identifiability of finite mixtures.
\newblock {\em The Annals of Mathematical Statistics}, pages 209--214.

\bibitem[Yao et~al., 2018]{Yao2018}
Yao, Y., Vehtari, A., Simpson, D., Gelman, A., et~al. (2018).
\newblock Using stacking to average {B}ayesian predictive distributions (with
  discussion).
\newblock {\em Bayesian Analysis}, 13(3):917--1003.

\bibitem[Yeh, 2011]{Yeh2011}
Yeh, C.-C. (2011).
\newblock H{\"o}lder's inequality and related inequalities in probability.
\newblock {\em International Journal of Artificial Life Research (IJALR)},
  2(1):54--61.

\bibitem[Zhong et~al., 2015]{Zhong2015}
Zhong, M., Goddard, N., and Sutton, C. (2015).
\newblock Latent bayesian melding for integrating individual and population
  models.
\newblock In {\em Advances in Neural Information Processing Systems}, pages
  3618--3626.

\end{thebibliography}

\newpage 

\appendix
\section{Appendix}
\renewcommand\thefigure{S\arabic{figure}}    
\setcounter{figure}{0} 

\subsection{Proofs}
\label{sec:appendix_proofs}

Here we provide a simple proof of Theorem~\ref{thm:normalisation} using H\"{o}lder's inequality.
\begin{proof}
We begin by noting that $\pi(\theta)$ can be re-written as:
\begin{equation}
\label{eq:pirewritten}
 \pi(\theta) \propto f_0(\theta)\prod_{j=1}^{K} \left(\frac{f_j(\theta)}{f_0(\theta)}\right)^{\alpha_j}.
\end{equation}
Let $X_j = \frac{f_j(\theta)}{f_0(\theta)}, j=1, 2,\ldots, K$. 
Then, integrating the expression in (\ref{eq:pirewritten}) is equivalent to finding 
\begin{equation}
\label{eq:expectations}
\mathbb{E}_{0}\left[\prod_{j=1}^KX_j^{\alpha_j}\right] \leq \prod_{j=1}^K \mathbb{E}_{0}[X_j]^{\alpha_j},
\end{equation}
where $\mathbb{E}_{0}[\cdot]$ is the expectation w.r.t $f_0$ and (\ref{eq:expectations}) follows from H\"{o}lder's inequality for expectations~\citep{Yeh2011}.
Since $\forall j$ we have $\mathbb{E}_{0}[X_j]^{\alpha_j} = \left(\int_{\boldsymbol\Theta}f_0(\theta)\frac{f_j(\theta)}{f_0(\theta)}\, d\theta\right)^{\alpha_j}=1^{\alpha_j}$, Theorem~\ref{thm:normalisation} is proven.
\end{proof}

To establish Theorem~\ref{thm:concavity}, we will need the following result from~\cite{Genest1984}.
\begin{lemma}
\label{lem:RPC_representation}
\textbf{Representation of a pooling operator with RPC}~\citep[eq. 3.1]{Genest1984}.
The \underline{only} relative propensity consistent operator can \underline{always} be represented by
\begin{equation}
 \label{eq:RPC_operator}
 \mathcal{T} \left( \boldsymbol F_\theta \right)(\theta) = \boldsymbol B\left( \boldsymbol F_\theta \right) c(\theta) \prod_{i=0}^K \left[f_i(\theta) \right]^{\alpha_i},
\end{equation}
with $\boldsymbol B\left( \boldsymbol F_\theta \right) > 0$, $c(\theta) >0$  and $\alpha_0, \alpha_1, \ldots, \alpha_K \geq 0$. 
\end{lemma}
We refer the reader to~\cite{Genest1984} for the proof.
In short, Lemma~\ref{lem:RPC_representation} is a uniqueness result; logarithmic pooling is the only operator with RPC (Remark~\ref{rmk:properties_RPC}) and every operator with RPC can be represented as in~(\ref{eq:RPC_operator}). 
Now we can state the proof of Theorem~\ref{thm:concavity}.
\begin{proof}
First, we will show by direct calculation that logarithmic pooling (LP) leads to a log-concave distribution.
Notice that each $f_i$ can be written as $ f_i(\theta) \propto e^{\nu_i(\theta)}$, where $\nu_i(\cdot)$ is a concave function.
We can then write
\begin{align*}
 \pi(\theta \mid \boldsymbol \alpha) &\propto \prod_{i=0}^{K} [\exp(\nu_i(\theta))]^{\alpha_i},\\
             &\propto \exp(\nu^{\ast}(\theta)),
\end{align*}
 where $\nu^{\ast}(\theta) = \sum_{i=0}^{K}\alpha_i\nu_i(\theta)$ is a concave function because it is a linear combination of concave functions.
 
We will now show that LP is the only operator that guarantees log-concavity when $\boldsymbol F_\theta$ is a set of concave distributions.
First, recall that LP is the only pooling operator that enjoys RPC (Remark~\ref{rmk:properties_RPC}).
With the goal of obtaining a contradiction, suppose that there exists a pooling operator $\mathcal{T}$ which is log-concave but does not enjoy RPC.
From Lemma~\ref{lem:RPC_representation}, we know that, under this assumption, $\mathcal{T}$ cannot be written in the form $\boldsymbol B(\boldsymbol F_\theta) c(\theta) \prod_{i=0}^K f_i(\theta)^{\alpha_i} $.
But every non-negative log-concave function $g(\theta)$ can be represented as
 \begin{equation}
 \label{eq:lc_rep}
  g(\theta) = a \cdot c(\theta) \cdot h(\theta),
 \end{equation}
with $a \geq 0$ and $c(\theta)$ and $h(\theta)$ non-negative and log-concave, but otherwise arbitrary.
Under the assumptions on $\boldsymbol F_\theta$, we have that $h(\theta) := \prod_{i=0}^K f_i(\theta)^{\alpha_i}$ is non-negative and log-concave.
The only restriction on $c(\theta)$ is that it be positive, it may very well be (log-)concave.
Therefore $\mathcal{T}$ can in fact be represented in the form of Lemma~\ref{lem:RPC_representation}, contradicting our initial assumption that $\mathcal{T}$ can at same time be log-concave but not enjoy RPC.  
\end{proof}

We now move on to the exponential family results in Section~\ref{sec:expofamily}.
To see that equation~(\ref{eq:entropypriorEF}) holds:
\begin{eqnarray*} 
H_\pi(\theta) & = & \mathbb{E}[-\log(\pi(\theta)], \\
              & = & - \int \log(\pi(\theta)) \pi(\theta) \, d\theta, \\
              & = & - \int (\log(K(a^*, b^*) + \theta a^* - s(\theta) b^*) \pi(\theta) \, d\theta, \\
              & = & - \log(K(a^*, b^*)) - a^*  \mathbb{E}[\theta] +  b^*  \mathbb{E}[s(\theta)].
\end{eqnarray*}
Likewise for equation~(\ref{eq:KLpriorEF}), we have 
\begin{eqnarray*} 
KL(f_i || \pi) & = & \mathbb{E}_\pi[\log(f_i(\theta)-\log(\pi(\theta)], \\
              & = & \int [\log( K(a_i,b_i) e^{\theta a_i - b_i s(\theta)}) - \log(K(a^*,b^*) e^{\theta a^* - b^* s(\theta)}) ] \pi(\theta) \, d\theta, \\
              & = & \int [\log( K(a_i,b_i)) - \log(K(a^*,b^*)) + (a_i - a^*) \theta  - (b_i - b^*) s(\theta) \pi(\theta) \, d\theta, \\
              & = & \log( K(a_i,b_i)) - \log(K(a^*,b^*)) + (a_i - a^*) \mathbb{E}_\pi[\theta] - (b_i - b^*) \mathbb{E}_\pi[s(\theta)]. 
\end{eqnarray*}

\newpage
\subsection{Computational details}
\label{sec:appendix_compdetails}

The analyses presented in this paper necessitated numerical optimisation to find the weights based on optimality criteria and Markov chain Monte Carlo (MCMC) to approximate posterior distributions.
All computations were carried out in the R~\citep{R2019} statistical computing environment, version 3.6.0. 
We provide implementations using the Stan~\citep{Carpenter2017} probabilistic programming language and R code for the methods, figures and tables presented in this paper can also be found at~\url{https://github.com/maxbiostat/opinion_pooling}.

\subsubsection{Optimisation procedures}
\label{sec:computation_opt}

In this section we give more detail on the optimisation procedures used to solve the problems in Sections~\ref{sec:maxent} and~\ref{sec:minKL}.
Since both problems are numerically unstable, we employ an strategy that starts the optimisation routine from $J = 1000$ overdispersed points in the unconstrained space, $\mathbb{R}^K$, obtained through the unit simplex transform~\citep{Betancourt2012}.
The procedure then picks the overall lowest/highest optimised value in order to avoid local minima/maxima.
We draw the initial values from a normal distribution with mean $0$ and variance $100^2$ and then employ the \verb|optim()| function to optimise the target functions (maximum entropy or minimum KL) using the L-BFGS algorithm~\citep{Byrd1995} with default settings.
We then choose the minimum/maximum achieved over the $J$ starting points in order to improve the chances of achieving a global optimum.

\subsubsection{Markov chain Monte Carlo \textit{via} Stan}
\label{sec:computation_mcmc}

Most of the posterior distributions discussed in this paper cannot be computed in closed-form and therefore we resort to Hamiltonian Monte Carlo~\citep{Neal2011}, in particular the No-U-Turn (NUTS) dynamic implementations available in Stan~\citep{Hoffman2014,Betancourt2017}.

For most computations, we employed four independent chains of $4000$ iterations each with the first $2000$ discarded as warm-up/burn-in.
Some models presented more challenging target distributions and we increased the number of iterations to $10 000$.
For all results reported, Monte Carlo error (MCSE) was well below 1\% the posterior standard deviation for all parameters, allowing for accurate computation of the relevant expectations.
In order to cope with challenging posterior geometry and ensure accurate computation, we used an target  acceptance probability of $0.99$ (\verb|adapt_delta = 0.99|, in Stan parlance) and up to $2^{15}$ leapfrog steps (\verb|max_treedepth = 15|).
All potential scale reduction factors ($\hat{R}$, \cite{Gelman1992}) were below $1.01$, indicating no convergence problems.

\subsubsection{Sampling-importance-resampling}
\label{sec:spIR}

In the context of Bayesian melding, while it is possible to employ HMC, it is sometimes preferable to employ custom algorithms that can better deal with the constraints imposed by the deterministic model.
In the original paper,~\cite[sec. 3.4]{Poole2000} propose a sampling-importance-resampling (SpIR) algorithm to sample from the posterior in~(\ref{eq:BMpoolposterior}), which we extend here to in order to accommodate varying weights.
\begin{enumerate}
\setcounter{enumi}{-1}
 \item Draw $k$ values from  $q_1(\theta)$, constructing $\boldsymbol \theta_k = (\theta^{(1)}, \theta^{(2)}, \ldots, \theta^{(k)} )$;
 \item Similarly, sample $\boldsymbol \alpha_k$ from $\pi(\boldsymbol \alpha)$;
 \item For each $\theta^{(i)} \in \boldsymbol\theta_k$ run the model to compute $\psi^{(i)} = M(\theta^{(i)})$, constructing $\boldsymbol \phi_k$;
 \item Obtain a density estimate of $q_1^\star(\phi)$ from  $\boldsymbol \phi_k$;
 \item Form the importance weights 
 \begin{equation}
 \label{eq:SpIRweights}
  w_i = t\left(\boldsymbol \alpha^{(i)}\right) \left(\frac{q_2(M(\theta^{(i)}))}{q_1^\star(M(\theta^{(i)}))}\right)^{1 - \boldsymbol \alpha^{(i)}} L_1(\theta^{(i)}) L_2(M(\theta^{(i)})),
 \end{equation}
where $t\left(\boldsymbol \alpha^{(i)}\right) = \left( \int_{\Phi} q_1^\ast(\phi)^{\alpha^{(i)}} q_2(\phi)^{1-\alpha^{(i)}} d\phi \right)^{-1}$ is computed using standard quadrature methods;
 \item (Re)Sample $l$ values from $\boldsymbol \theta_k$ according to the weights $\boldsymbol w_k$.
\end{enumerate}
The quadrature-based normalisation in step in 4 can be replaced with an importance sampling or MCMC estimate when the dimension of either $\phi$ or $\theta$ is large, but this is not explored here.

\newpage
\subsection{Bowhead population growth model: details}
\label{sec:appendix_bowhead}

For convenience, here we will describe the priors and likelihoods used by~\cite{Poole2000} in their analysis of the bowhead whale population model, as well as some of our modelling choices.
For $P_0$ only a shifted gamma prior is available, i.e.,
\begin{equation*}
 q(P_0) = \frac{b_{P_0}^{a_{P_0}}}{\Gamma(a_{P_0})} (P_0 - s_{P_0})^{a_{P_0}-1} \exp\left(-b_{P_0}(P_0 - s_{P_0})\right), \: P_0 > s_{P_0},
\end{equation*}
with $s_{P_0} = 6400$, $a_{P_0} = 2.8085$ and $b_{P_0} = 0.0002886$.
The maximum sustainable yield rate (MSYR) is assigned Gamma prior with parameters $a_{\text{MSYR}} = 8.2$ and $a_{\text{MSYR}} = 372.7$.

The size of the bowhead population in 1993, $P_{1993}$, is an output of the model for which there are both a prior and a likelihood.
The prior ($q_2$) is a Gaussian distribution with mean $\mu_{1993} = 7800$ and standard deviation $\sigma_{1993} = 1300$, while the likelihood ($L_2$) is also a Gaussian distribution but with mean $\mu_{1993}^\prime = 8293$ and standard deviation $\sigma_{1993}^\prime = 626$.
For the rate of increase (ROI),~\cite{Poole2000} use a likelihood that is proportional to  $\exp(a + b \times t_8) - 1$, with $a = 0.0302$ and $b = 0.0068$ where $t_8$ is a random variable with Student t distribution with $\nu = 8$ degrees of freedom.
This leads to the density
\begin{equation*}
 L(\text{ROI} \mid \nu, a, b) = \frac{\Gamma((\nu + 1)/2)}{\Gamma(\nu/2)}\frac{1}{\sqrt{\nu\pi}} \left( 1 + \left((\log(\text{ROI} + 1) - a)/b\right)^2 \right)^{-(\nu+ 1)/2}\frac{1}{|b(\text{ROI} + 1)|},\: \text{ROI} > -1.
\end{equation*}

\begin{figure}[!ht]
\begin{center}
\includegraphics[scale=.5]{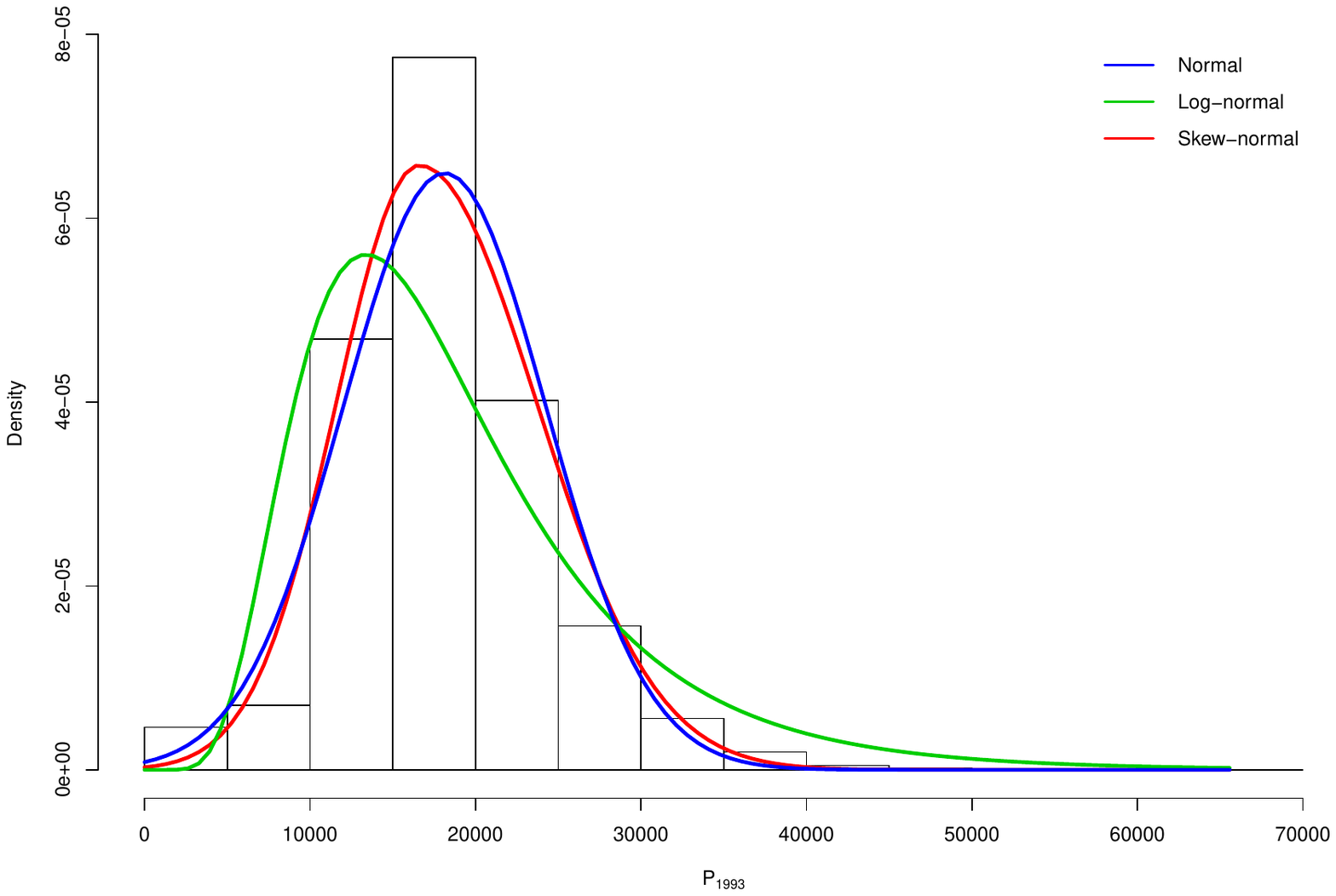}
\end{center}
\caption{\textbf{Induced distribution on $P_1993$ ($q_1^\ast$) and approximating distributions}.
We present the histogram of $100, 000$ simulations from the prior.
Lines show the densities of the three distributions considered.
}
\label{sfig:induced_P1993}
\end{figure}

For the Stan implementation, we approximate the distribution induced on $P_{1993}$ by the prior on $P_0$ and $\text{MSYR}$ and transformation in~(\ref{eq:popmodel_bowhead}), $q_1^\ast$,  by a normal distribution with mean $\mu_{\text{ind}} = 18137.70$ and standard deviation $\sigma_{\text{ind}} = 6146.85$ .
This step deserves a bit more consideration.
As discussed by~\cite{Poole2000}, $q_1^\ast$ is very diffuse and likely has heavier tails than a normal distribution.
Hence it would make sense also to consider the skew-normal and log-normal families as approximating distributions.
On the other hand, we note that approximating $q_1^\ast$ with a normal distribution allows closed-form computation of the coherised prior $\tilde{q}_{\Phi}(P_{1993})$.
In Figure~\ref{sfig:induced_P1993} we show the densities of a normal, skew-normal and log-normal distributions fitted to $100, 000$ simulations from $q_1^\ast$ by maximum likelihood.
While the skew-normal provides better fit (AIC: 1150814) we do not feel the difference in fit to the normal (AIC: 1152288) justifies the increased technical overhead of not being able to compute the coherised prior in closed-form.
Both distributions provide much superior fit than the log-normal (AIC: 1181753).

Note that the sampling-importance-resampling discussed in Section~\ref{sec:spIR} does not necessitate any parametric approximation, employing a density estimation method instead.

\begin{figure}[!ht]
\begin{center}
\includegraphics[scale=.65]{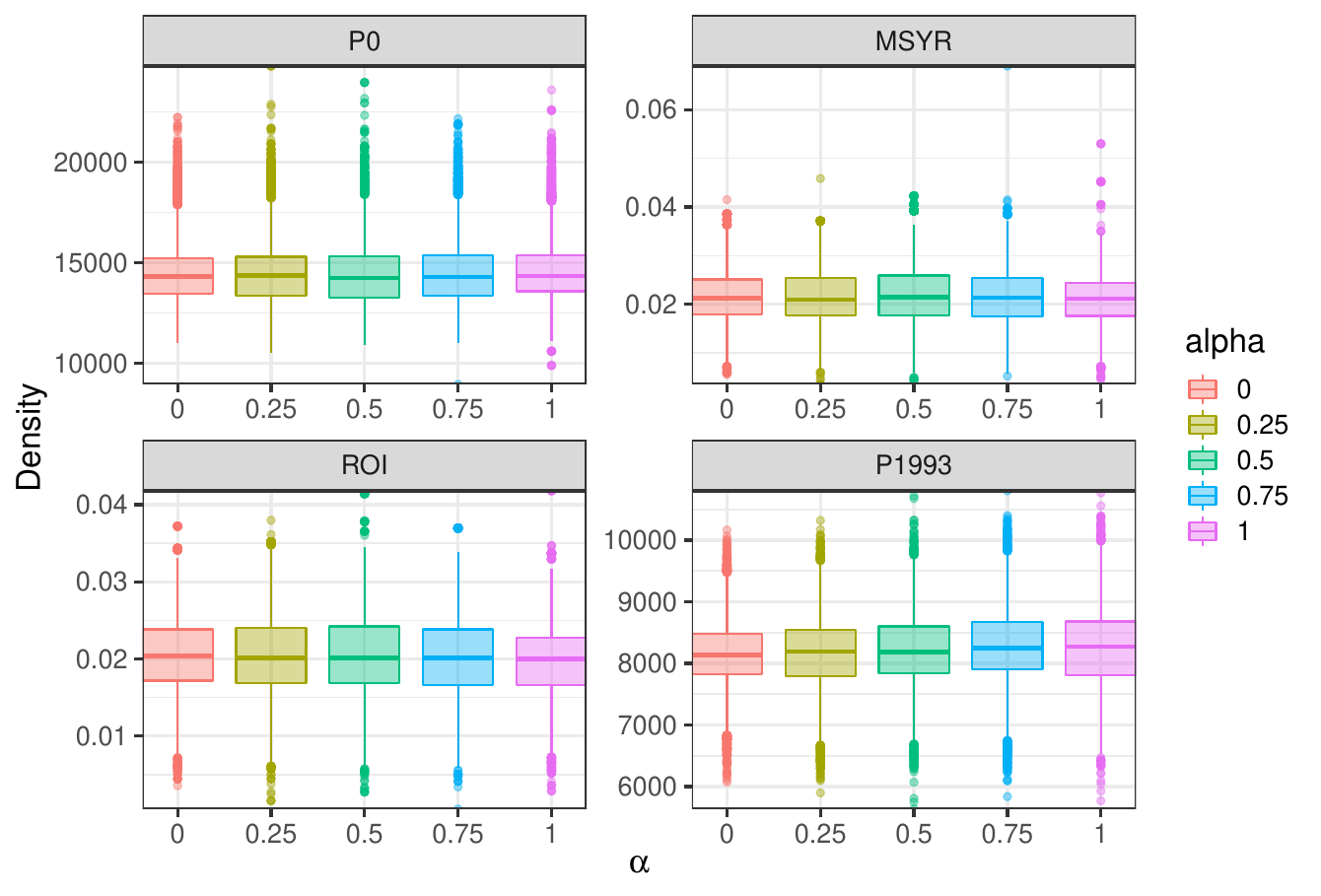}
\end{center}
\caption{\textbf{Sensitivity of posterior inferences to varying the value of $\alpha$, bowhead population model}.
}
We show the posterior distributions obtained by SpIR for $P_0$, MSYR, ROI and $P_1993$ as we fix $\alpha$ to different values.
\label{sfig:alpha_sensitivity_bowhead}
\end{figure}

In Figure~\ref{sfig:alpha_sensitivity_bowhead} we show the posterior distributions obtained with different values of $\alpha$ using SpIR.

\newpage
\subsection{Pooling of common distributions}
\label{sec:appendix_common_poolings}

In the main text we give the pooled distributions if one assumes a set of Beta (Section~\ref{sec:survivalProbs}) or Gaussian (Section~\ref{sec:metaAnalysis}) distributions as the expert opinions.
In this section we give further results for the pooling of commonly used distributions.

\subsubsection{Gamma}
\label{sec:gamma}
Suppose $K + 1$ experts are called upon to elicit prior distributions for a quantity $\lambda \in \mathbb{R}^+$.
A convenient parametric choice for $\mathbf{F}_\lambda$ is the Gamma family of distributions, for which densities are of the form
$$ f_i(\lambda;a_i,b_i) = \frac{b_i^{a_i}}{\Gamma(a_i)} \lambda^{a_i-1} e^{-b_i\lambda}.$$
The log-pooled prior $\pi(\lambda)$ is then
\begin{align}
\nonumber
\pi(\lambda)&= t(\boldsymbol\alpha)\prod_{i=0}^{K}f_i(\lambda;a_i,b_i)^{\alpha_i},\\
\nonumber
&\propto \prod_{i=0}^{K} \left(\lambda^{a_i-1} e^{-b_i\lambda}\right)^{\alpha_i},\\
\label{eq:gammapois}
&\propto \lambda^{a^*-1} e^{-b^*\lambda},
\end{align}
where $a^* =\sum_{i=0}^{K}\alpha_ia_i$ and $b^* = \sum_{i=0}^{K}\alpha_ib_i$.
Noticing (\ref{eq:gammapois}) is the kernel of a gamma distribution with parameters $a^*$ and $b^*$, $H_{\pi}(\lambda)$ becomes
\begin{equation}
\label{eq:entropygamma}
H_{\pi}(\lambda; \boldsymbol\alpha) = a^* - \log b^* + \log \Gamma(a^*) + (1-a^*)\psi(a^*),
\end{equation}
where $\psi(\cdot)$ is the digamma function.
The Kullback-Leibler divergence between the pooled density $\pi$ and each density is:
\begin{equation}
 \label{eq:KLgamma}
 \text{KL}(\pi || f_i) = (a_i-a^*)\psi(a_i) - \log\Gamma(a_i) + \log\Gamma(a^*) + a^*\left(\log\frac{b_i}{b^*}\right) + \frac{a_i}{b_i}(b^*-b_i).
\end{equation}

\subsubsection{Log-normal} 
\label{sec:log-normal}
Another popular choice for modelling a quantity $\eta \in \mathbb{R}^+$ is the log-normal family.
Following the results given in Section~\ref{sec:metaAnalysis}, we know that the pool of log-normal distributions with parameters $\mu_i$ and $\sigma_i^2$ is a log-normal distribution with parameters $\mu^\star = \frac{\sum_{i=0}^K w_im_i}{\sum_{i=0}^K w_i}$ and ${\sigma^\star}^2 = [\sum_{i=0}^K w_i]^{-1}$,  where $w_i = \alpha_i/\sigma_i^2$.

The entropy function is then:
\begin{align}
 \nonumber
 H_{\pi}(\eta ; \boldsymbol\alpha) &= \log_2(e)\log\left(\sigma^\star\exp\left(\mu^\star + \frac{1}{2}\right) \sqrt{2\pi}\right),\\
 \label{eq:log-normalpoolentropy}
 & = \log_2(e) \left[\log\left(\sigma^\star\right) +  \mu^\star + \frac{1}{2} + \log(\sqrt{2\pi})\right].
\end{align}

The KL divergence evaluates to
\begin{equation}
 \label{eq:KL_log-normal}
 \text{KL}(\pi || f_i) = \frac{1}{2\sigma_i^2} \left[ \left( \mu^\star - \mu_i \right)^2 + {\sigma^\star}^2 - \sigma_i^2 \right] + \log\left(\frac{\sigma_i^2}{{\sigma^\star}^2}\right).
\end{equation}

\subsubsection{Poisson}
\label{sec:poisson}

If the quantity of interest is a count $y = 0, 1, \ldots, $ and $\boldsymbol F_y$ is a set of Poisson distributions with rate parameters $\boldsymbol\lambda = \{\lambda_0, \lambda_1, \ldots, \lambda_K \}$.
We have 
\begin{align}
\nonumber
 \pi(y) &\propto \prod_{i = 0}^K \left(\frac{\lambda_i^y}{y!}\right)^{\alpha_i},\\
 \label{eq:pooled_Poisson}
 \pi(y) &= \frac{\exp(-\lambda^\star) {\lambda^\star}^{y}}{y!},\: \text{with}\: \lambda^\star = \prod_{i = 0}^K \lambda_i^{\alpha_i}.
 \end{align}
The entropy of the pooled distribution is 
\begin{equation}
 \label{eq:entropy_Poisson}
 H_\pi(y; \boldsymbol\alpha) = -\lambda^\star \log\left(\frac{\lambda^\star}{e}\right) + E_\pi\left[ \log(k!) \right],
\end{equation}
where the latter term cannot be evaluated in closed-form, but efficient approximations exist~\citep{Evans1988}.

The KL divergence is
\begin{align}
 \nonumber 
 \text{KL}(\pi || f_i) &= \lambda^\star \log\left( \frac{\lambda^\star}{\lambda_i} \right) + \lambda_i - \lambda^\star,\\
 \label{eq:KL_Poisson}
 & = \lambda^\star \left[ \sum_{k = 0}^K \alpha_k \log(\lambda_k) - \log(\lambda_i) \right] + \lambda_i - \lambda^\star.
\end{align}
\newpage
\subsection{Supplementary figures}

\begin{figure}[!ht]
\begin{center}
\subfigure[][Ratios of marginal likelihoods]{\includegraphics[scale=.45]{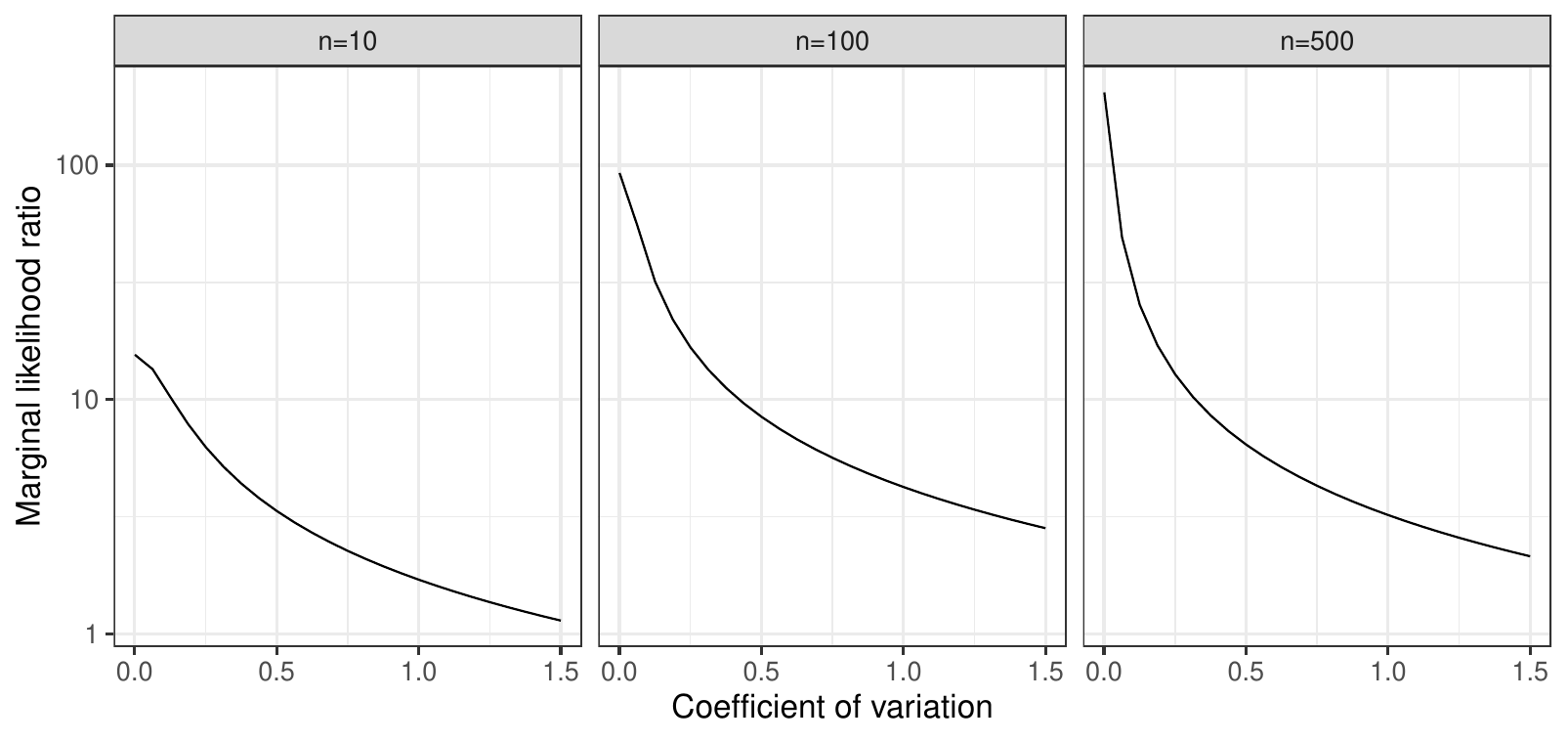}}
\subfigure[][Ratios of posterior weights]{\includegraphics[scale=.45]{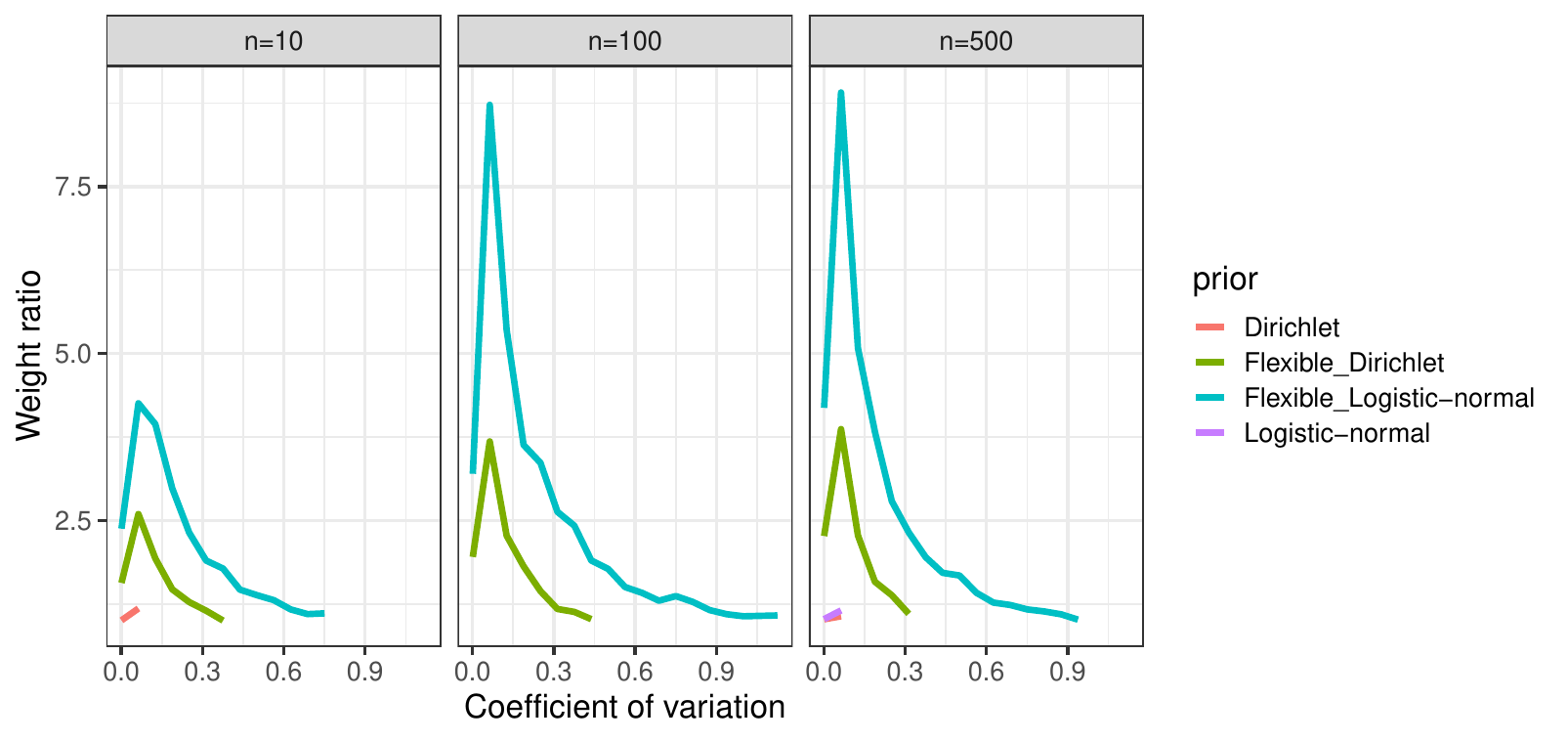}}
\end{center}
\caption{\textbf{Marginal likelihood and weight ratios for the simulated situation with one correct expert, various strengths of evidence, Gaussian example}.
Panel (a) shows the ratio between the largest and second largest marginal likelihoods ($r_l$) as the correct expert's coefficient of variation ($c_2$) changes, while panel (b) shows the ratio between the largest and second largest posterior mean weights ($r_w$) in the same settings.
Vertical tiles show the observed data and colours in panel (b) show the hyperprior on $\boldsymbol\alpha$.
``Flexible'' priors are a Dirichlet(1/10, 1/10, 1/10, 1/10, 1/10) and the corresponding moment-matching logistic-normal.
We interrupt the lines for values of $c_2$ for which expert 2, the correct expert, does not attain the largest posterior weight (see Section~\ref{sec:learning_rate_Gaussian}). 
}
\label{fig:one_correct_results_normal}
\end{figure}

\begin{figure}[!ht]
\begin{center}
\subfigure[][Marginal posterior of $\alpha_0$]{\includegraphics[scale=.45]{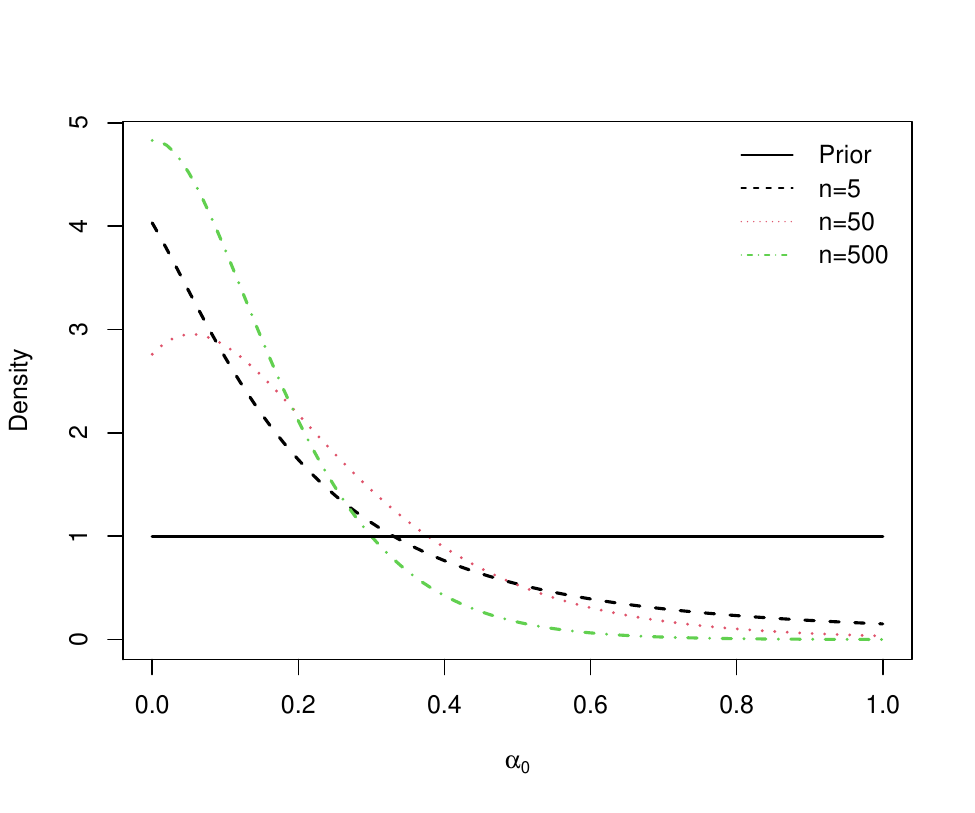}}
\subfigure[][Posterior concentration, uniform prior]{\includegraphics[scale=.55]{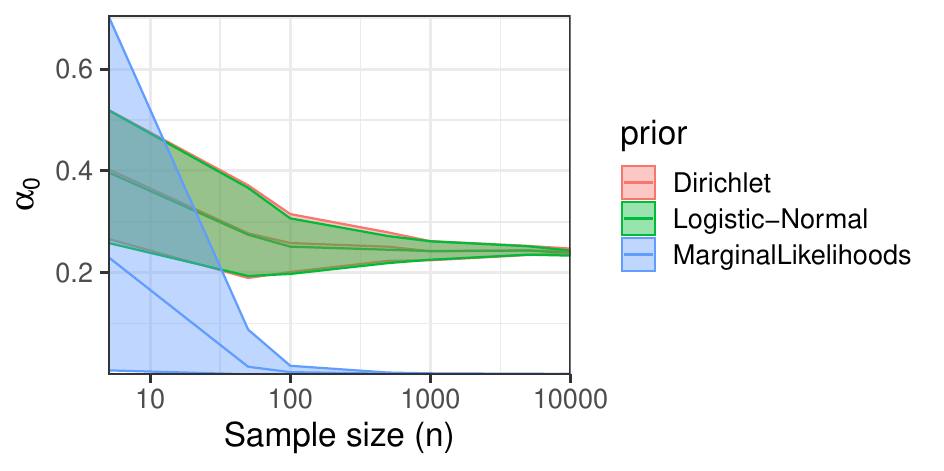}}\\
\subfigure[][Posterior concentration, $\operatorname{Beta}(1/10, 1/10)$ prior]{\includegraphics[scale=.55]{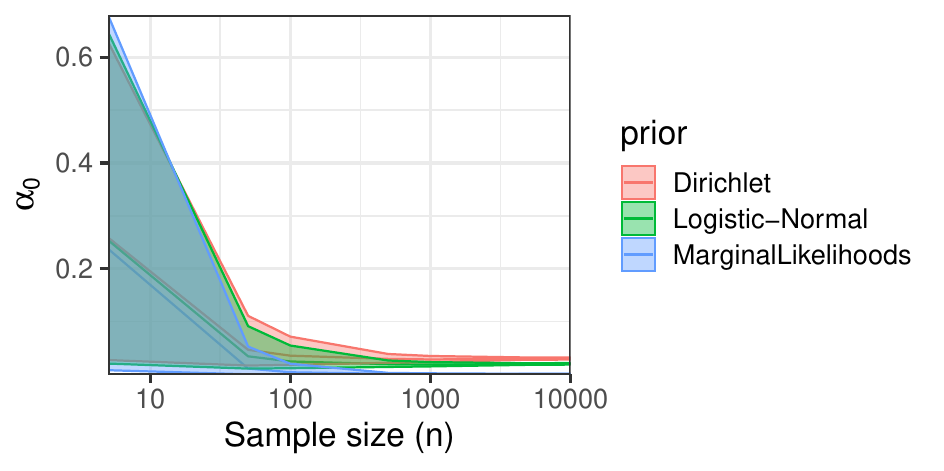}}
\end{center}
\caption{\textbf{Posterior concentration for the weights in a two-expert setting, Gaussian conjugate analysis with known variance ($\sigma^2$)}.
In this experiment we set the expert hyperparameters to $m_0 = 1$ $v_0 = (1/4)^2$, $m_1 = 2$, $v_1 = (1/2)^2$ and the true data-generating parameters at $\mu = 2$ and $\sigma^2 = 1^2$.
In panel (a) we show the marginal posterior of $\alpha_0$ for various sample sizes under a Beta prior for $\alpha_0$ with parameters $a = b = 1$.
Panel (b) shows the posterior mean of $\alpha_0$ versus sample size under repeated sampling, using 100 simulated data sets per sample size. 
Bands correspond to the 95\% quantiles of the sampling distribution of the posterior mean.
In this experiment the prior on $\alpha_0$ is a uniform prior over $(0, 1)$.
For comparison we show the BMA weight for expert 0, computed from the marginal likelihoods.
In panel (c) we show the same experiment, but using the ``flexible'' prior discussed in the main text, corresponding to a $\operatorname{Beta}(1/10, 1/10)$ prior on $\alpha_0$.
}
\label{fig:concentration_results_normal}
\end{figure}
\end{document}